%% file: RT_paper.tex
\documentclass[preprint]{elsarticle}
\usepackage[dvips]{graphicx}
\usepackage{color,amsmath,latexsym,amssymb,enumerate}
\usepackage{subfigure}

\journal{Journal of Computational Physics}

\input{latexdefs.tex}

\begin{document}
\begin{frontmatter}

\title{Self-Consistent Solution of Cosmological
  Radiation-Hydrodynamics and Chemical Ionization}

\author[smuadd]{Daniel R. Reynolds\corref{cor1}}
\ead{reynolds@smu.edu}
\author[llnladd]{John C. Hayes}
\ead{jchayes@llnl.gov}
\author[cassadd]{Pascal Paschos}
\ead{ppaschos@minbari.ucsd.edu}
\author[cassadd,physadd]{Michael L. Norman}
\ead{mlnorman@ucsd.edu}

\address[smuadd]{Mathematics, Southern Methodist University,
  Dallas, TX 75275-0156} 

\address[llnladd]{Lawrence Livermore National Lab, PO Box 808, L-551,
  Livermore, CA 94551}

\address[cassadd]{Ctr. for Astrophysics and Space Sciences,
  U.C.~San Diego, La Jolla, CA 92093}

\address[physadd]{Physics Department,
  U.C.~San Diego, La Jolla, CA 92093}

\begin{abstract}
  We consider a PDE system comprising compressible hydrodynamics,
  flux-limited diffusion radiation transport and chemical ionization
  kinetics in a cosmologically-expanding universe.  Under an
  operator-split framework, the cosmological hydrodynamics
  equations are solved through the Piecewise Parabolic Method, as
  implemented in the Enzo community hydrodynamics code.  The remainder
  of the model, including radiation transport, chemical ionization
  kinetics, and gas energy feedback, form a stiff coupled PDE system,
  which we solve using a fully-implicit inexact Newton approach, and
  which forms the crux of this paper.  The inner linear Newton systems
  are solved using a Schur complement formulation, and employ a
  multigrid-preconditioned conjugate gradient solver for the inner
  Schur systems.  We describe this approach and provide results on a
  suite of test problems, demonstrating its accuracy, robustness, and
  scalability to very large problems.
\end{abstract}

\end{frontmatter}


\input{introduction.tex}

\input{model.tex}
\input{solution.tex}
\input{results.tex}
\input{conclusions.tex}
\input{acknowledgements.tex}


\bibliography{sources}
\bibliographystyle{elsarticle-num}

\end{document}

%% file: latexdefs.tex
\renewcommand{\(}{\left(}
\renewcommand{\)}{\right)}
\newcommand{\vb}{{\bf v}_b}
\newcommand{\xvec}{{\bf x}}

\newcommand{\rhob}{\rho_b}
\newcommand{\dt}{\Delta t}
\newcommand{\Enu}{E_{\nu}}

\newcommand{\eh}{e_h}
\newcommand{\ec}{e_c}

\newcommand{\mn}{{\tt n}}

\newcommand{\mL}{{\mathcal L}}
\newcommand{\mD}{{\mathcal D}}

\newcommand{\mDE}{\mD_{E}}

\newcommand{\mLE}{{\mathcal L}_{E}}

\newcommand{\mLe}{\mL_e}

\newcommand{\mLn}{\mL_{\mn_i}}
\newcommand{\mS}{\mathcal S}

%% file: introduction.tex
\section{Introduction}
\label{sec:introduction}

A fundamental physics component in cosmology and astrophysics
applications is the transport of ionizing radiation along with the
interaction of that radiation with the hydrodynamic motion and
ionization state of the surrounding gas. One example currently receiving a great
deal of attention is cosmic reionization \cite{BarkanaLoeb2007}, our
motivation in this work. Observations indicate that an early
population of UV emitting galaxies photoionized the intergalactic
hydrogen and helium gas when the universe was about 800 million years
old (redshift $\sim 8$). A computational challenge is to calculate
this process self-consistently, coupling the radiative transfer of
ionizing photons, the ionization kinetics and photo-heating of the
gas, and the attendant hydrodynamic motions. This problem is
challenging because the physics is numerically stiff and cosmic
reionization is intrinsically three-dimensional, involving the growth,
percolation, and overlap of ionization zones around an irregular and
evolving distribution of galaxies with time-dependent luminosities. In
addition, the problem inherits the large range of spatial scales
($\sim 10^5$) intrinsic in galaxy formation simulations, necessitating
the use of spatially adaptive mesh or particle--tree methods and
large-scale parallel computing \cite{NormanEtAl2007,Springel2005}. 

A variety of 3D radiative transfer methods are under development to
tackle this problem \cite{IlievEtAl2006}. These necessarily simplify
the description of the radiation field in order to render the problem
tractable. These methods include ray tracing using long and short
characteristics
\cite{MellemaEtAl2006,SusaUmemura2004,WhalenNorman2006,RijkhorstEtAl2006},
Monte Carlo \cite{MaselliEtAl2003,SemelinEtAl2007}, and moment methods  
\cite{GnedinAbel2001,HayesNorman2003,PaschosEtAl2007}. However, only
some of these codes allow solution of the coupled problem on
spatially-adaptive grids, and very few allow distributed-memory
parallelism. More importantly, all of these codes handle the
interactions between hydrodynamic, radiative, and chemical processes
in an explicit, operator-split fashion, which ignores stiff couplings
that often arise between these components. When this happens, such codes must dramatically
reduce allowable timesteps, or resort to subcycling, to ensure
stability and accuracy of the coupled simulations.  

Radiation transport and chemical ionization time scales are much
faster than typical hydrodynamic time scales. This is particularly
evident in dense gas bound to galaxies where recombination and light
crossing times are short compared to the age of the universe (Hubble
time). Moreover, these processes are very tightly coupled since
radiation induces ionization that in turn affects opacities. While
time-explicit, adaptive subcycling schemes have been developed that
are capable of returning accurate solutions in all regimes of interest
\cite{WhalenNorman2006}, it is our view that for both computational
efficiency and solution accuracy, tightly-coupled implicit methods
require investigation. Here we present such a method.  

We solve ionizing radiation transport, chemical ionization kinetics,
and gas photo-heating using a fully implicit inexact Newton
method. This algorithm is coupled to a cosmological hydrodynamics code
through an explicit, operator-split formalism. The inner linear Newton
systems are solved using a Schur complement formulation, which neatly
decouples the local microphysics from the transport
calculation. Radiation transport is modeled in the flux-limited
diffusion approximation for simplicity, although our approach can be
easily generalized to higher-order moment schemes. The use of Schur
complements allows the application of optimal and scalable multigrid
methods for the solution of the scalar radiation diffusion
equation. We describe our algorithm in detail. We then illustrate our
method's accuracy, robustness, and parallel scalability against a
suite of verification tests of increasing size and complexity. In its
current implementation, we are restricted to calculations on uniform
Cartesian grids. An extension of our algorithm on block structured
adaptive meshes is under development, and requires only modifications
to the inner multigrid linear solver.  

The paper is organized as follows. In Sec. 2 the governing equations
for cosmological radiation hydrodynamics are presented under two
different assumptions about the radiation--matter coupling: a two
temperature model assuming local thermodynamic equilibrium (LTE), and
a non-LTE ionization kinetics multispecies model. In Sec.~3 we
describe our solution procedures for splitting off the hydrodynamic
calculation, and our coupled implicit radiation-ionization-gas energy
system. Results from solution verification tests are presented in
Sec.~4, as well as parallel scalability tests. Conclusions follow in
Sec.~5.

%% file: model.tex
\section{Cosmological Radiation-Hydrodynamics-Ionization Model}
\label{sec:model}

We consider the coupled system of partial differential equations
\begin{align}
  \label{eq:cons_mass}
  \partial_t \rhob + \frac1a \vb \cdot \nabla
    \rhob &= -\frac1a \rhob \nabla\cdot\vb, \\
  \label{eq:cons_momentum}
  \partial_t \vb + \frac1a\(\vb\cdot\nabla\)\vb &=
    -\frac{\dot{a}}{a}\vb - \frac{1}{a\rhob}\nabla p - \frac1a
    \nabla\phi, \\
  \label{eq:cons_energy}
  \partial_t e + \frac1a\vb\cdot\nabla e &=
    - \frac{2\dot{a}}{a}e
    - \frac{1}{a\rhob}\nabla\cdot\left(p\vb\right) 
    - \frac1a\vb\cdot\nabla\phi + G - \Lambda \\
  \label{eq:chemical_ionization}
  \partial_t \mn_i + \frac{1}{a}\nabla\cdot\(\mn_i\vb\) &=
    \alpha_{i,j} \mn_e \mn_j - \mn_i \Gamma_{i}^{ph}, \qquad
    i=1,\ldots,N_s \\
  \label{eq:cons_radiation}
  \partial_t E + \frac1a \nabla\cdot\(E \vb\) &= 
    \nabla\cdot\(D\nabla E\) - m\frac{\dot{a}}{a}E + 4\pi\eta - c \kappa E.
\end{align}
Here, the first three equations
\eqref{eq:cons_mass}-\eqref{eq:cons_energy} correspond to the
equations of ideal gas dynamics in a coordinate system that is
comoving with the expanding universe \cite{BryanEtAl1995}.  These
equations correspond to mass, momentum and energy conservation,
respectively, in which $\vb\equiv a(t)\dot{\xvec}$ is the proper
peculiar baryonic velocity, $p$ is the proper pressure, and the total
gas energy per unit mass is given by $e$.  The modified gravitational
potential is given by $\phi$, which satisfies the comoving form of
Poisson's equation, 
\begin{equation}
\label{eq:gravity}
  \nabla^2\phi = \frac{4\pi g}{a}(\rhob + \rho_{dm} - \langle \rho \rangle),
\end{equation}
where $g$ provides the gravitational constant, $\rhob$ and $\rho_{dm}$
are the baryonic and dark matter densities, respectively, and 
$\langle \rho \rangle$ is the cosmic mean density.
The densities $\rho_i$ are comoving, relating to the proper densities 
through the relation $\rho_i \equiv \rho_{i,\text{proper}}a(t)^3$.
Here $a(t)\equiv(1+z)^{-1}$ denotes the cosmological expansion
parameter for a smooth homogeneous background, where the redshift $z$
is a function of time only.  All spatial derivatives are taken with
respect to the comoving position $\xvec\equiv{\bf r}/a(t)$. 
The hydrodynamics equations are closed as usual with the ideal gas
equation of state, 
\begin{equation}
\label{eq:eos}
  e = \frac{p}{\rhob(\gamma-1)} + \frac12|\vb|^2,
\end{equation}
where $\gamma$ is the ratio of specific heats, taken to be $5/3$
throughout this work.

The hydrodynamics equations are coupled with the elemental species
rate equations \eqref{eq:chemical_ionization} and an equation
describing the flux-limited diffusion (FLD) approximation of radiation 
transport in a cosmological medium \eqref{eq:cons_radiation}
\cite{HayesNorman2003,Paschos2005}.  In these equations $i$ denotes
the $i^{th}$ chemical species (out of $N_s$ total), $\mn_i$ is the
comoving number density, $\mn_e$ is the electron number density,
$\mn_j$ corresponds to other ions that react with the species $i$, and
$\alpha_{i,j}$ are the rate coefficients defining these
interactions \cite{AbelEtAl1997,HuiGnedin1997}; $E$ corresponds to the
comoving radiation energy density.  The parameter $m$ controls whether
$E$ is monochromatic at a specified frequency $\nu$ ($m=0$), or an
integrated grey radiation energy density ($m=1$).

The baryonic gas is coupled to collisionless dark matter solely through
their self--consistent gravitational field via \eqref{eq:gravity}. 
The dark matter density is evolved using the Particle-Mesh method
described in
\cite{HockneyEastwood1988,NormanBryan1999,OSheaEtAl2004}. As the
N-body method is standard and not the focus of this paper, we do not
elaborate on it here.

\subsection{Model Coupling}
\label{subsec:equation_coupling}

In addition to the advective components of the chemistry and radiation
equations, coupling between these equations arise through a number of
spatially-local reaction terms.  The radiation energy density and
chemical number densities affect the gas energy through the heating
and cooling rates $G$ and $\Lambda$, respectively.  The ionization
and recombination rates $\alpha_{i,j}$ and emissivity $\eta$ depend on
the gas temperature, 
\begin{equation}
\label{eq:temperature}
  T = (\gamma-1)\frac{p\,\mu\, m_p}{\rhob\, k_b},
\end{equation}
where $m_p$ corresponds to the mass of a proton, $\mu$ corresponds to
the local molecular weight, and $k_b$ is Boltzmann's constant.
Finally, the photoionization rate $\Gamma_i^{ph}$ depends on the
radiation energy density, and the opacity $\kappa$ depends on the
state of chemical ionization.

In determining these coupling terms we distinguish between two cases:
those in local thermodynamic equilibrium (LTE) and those that are not
(nLTE).  In the LTE case the chemical species are assumed to be in
equilibrium, and hence their equations \eqref{eq:chemical_ionization}
may be omitted from the time-dependent system
\eqref{eq:cons_mass}-\eqref{eq:cons_radiation}.  For problems in this
regime, the coupling terms resemble those typically encountered in
radiation-hydrodynamics simulations
\cite{HayesNorman2003,HayesEtAl2006,HowellGreenough2003}, 
\begin{equation}
\label{eq:LTE_coupling}
\begin{split}
  \eta_{LTE}(T) &= \kappa_P B = \frac{\kappa_P \sigma_{SB}}{\pi} T^4, \\
  G_{LTE}(\rhob,E) &= \frac{c \kappa}{\rhob} E, \\
  \Lambda_{LTE}(\rhob,T) &= \frac{4\pi}{\rhob} \eta_{LTE}(T),
\end{split}
\end{equation}
where $c$ is the speed of light, $\sigma_{SB}$ is the Stefan-Boltzmann
constant, and $\kappa_P$ and $\kappa$ correspond to the
problem-dependent Planck mean and total opacities for the gas.

For simulations that may not be approximated as being in local thermodynamic
equilibrium, these coupling terms involve the dynamically-changing
chemical ionization states.  Here, the combined opacity depends on the
local ionization states $\mn_i$, the emissivity
$\eta$ depends on both $T$ and $\mn_i$, the gas heating rate $G$
depends on both $E$ and $\mn_i$,  
and the gas cooling rate $\Lambda$ depends on $T$ and $\mn_i$, with
the corresponding formulas given in the references
\cite{AbelEtAl1997,Black1981,Osterbrock1989,Cen1992,RazoumovEtAl2002}.

\subsection{Cosmological Flux-Limited Radiation Diffusion Model Details}
\label{subsec:fld_details}

We derive the cosmological flux-limited radiation diffusion equation 
\eqref{eq:cons_radiation} from the general multi-frequency version
\cite{Paschos2005}, 
\begin{equation}
\label{eq:cons_radiation_multifreq}
     \partial_{t} \Enu + \frac1a \nabla\cdot(\Enu\vb) =
     \nabla\cdot(D\,\nabla\Enu) + \nu \frac{\dot{a}}{a}\partial_{\nu}\Enu 
     + 4\pi\eta_{\nu} - c \kappa_{\nu} \Enu.
\end{equation}
Through assumption of a given radiation frequency spectrum,
$\chi_E(\nu)$, the frequency-dependent radiation energy density
may be written in the form
$\Enu(\xvec,t,\nu)=\tilde{E}(\xvec,t)\,\chi_E(\nu)$.  With this, we
define the single ``grey'' radiation energy density used in the
model \eqref{eq:cons_mass}-\eqref{eq:cons_radiation} as
\begin{equation}
\label{eq:grey_definition}
  E(\xvec,t) = \int_{\nu_0}^{\infty} \Enu(\xvec,t,\nu) d\nu =
  \tilde{E}(\xvec,t) \int_{\nu_0}^{\infty} \chi_E(\nu)d\nu.
\end{equation}
The radiation equation \eqref{eq:cons_radiation} may then be derived
through integration of the equation
\eqref{eq:cons_radiation_multifreq} over frequencies ranging from the 
ionization threshold of Hydrogen ($h \nu_0=13.6$ eV) to infinity;
integration of the term $\nu \frac{\dot{a}}{a}\partial_{\nu}\Enu$ 
gives rise to the term $-\frac{\dot{a}}{a}$ in \eqref{eq:cons_radiation}.
We note that this approximation \eqref{eq:grey_definition} is valid
only if the assumed spectrum $\chi_E(\nu)$ is defined such that the
indefinite integral exists, as is the case for quasar and
stellar type spectra where it scales with frequency as $E_{\nu} \propto
\nu^{-\beta_{q}}$ where $\beta_{q} > 1$.  However, through this
formulation we may also consider problems involving a monochromatic
radiation energy density, since such energy densities may also be
expanded as $\Enu(\xvec,t,\nu)=\tilde{E}(\xvec,t)\,\chi_E(\nu)$, where for
radiation at the monochromatic frequency $\nu_k$, the assumed spectrum
is given through the Dirac-delta function $\chi_E(\nu)=\delta_{\nu_k}(\nu)$.
In such cases, the term $\nu \frac{\dot{a}}{a}\partial_{\nu}\Enu$
vanishes, giving rise to the parameter $m$ in \eqref{eq:cons_radiation}.  
For standard grey radiation approximations, we assume a radiation
spectrum of the form of either a power law, $\chi_E(\nu) =
\left(\frac{\nu}{\nu_0}\right)^{\alpha}, \alpha<-1$, or as a $T_B$
blackbody spectrum, $\chi_E(\nu) = 8\pi h\left(\frac{\nu}{c}\right)^3
\big/\left(\exp\left(\frac{h\nu}{k_bT_B}\right)-1\right)$. 

As is standard with FLD approximations to radiation transfer, one must
take special care in construction of the diffusion coefficient
function $D$.  In its simplest form, $D$ may be written as 
$D = \frac{c}{3\kappa_T}$, 
where $\kappa_T=\kappa_A+\kappa_S$ is the {\em total extinction coefficient}, 
$\kappa_A$ corresponds to total absorption (here taken to be the
opacity $\kappa$) and $\kappa_S$ corresponds to scattering
\cite{HayesEtAl2006}.  Use of this form for the diffusion 
coefficient, however, results in an infinite signal speed for the
radiative flux $-\frac1a D\,\nabla E$.  To preserve causality, the
analytic form of $D$ is modified with a dimensionless 
{\em flux-limiter} whose particular form may be tuned to the specific 
problem of interest, but whose overriding purpose is to guarantee that
the radiation transfer equation \eqref{eq:cons_radiation} gives the
correct numerical behavior in the limiting cases of (nearly) isotropic
and free-streaming radiation.  Several choices for flux-limited forms
of $D$ have been proposed in the literature 
\cite{Levermore1984,LevermorePomraning1981}.  We consider the
diffusion coefficient to be of the form 
\begin{equation}
\label{eq:limiter}
\begin{split}
   D(E) &= \left[\begin{array}{ccc} 
       D_1(E) & 0 & 0 \\
       0 & D_2(E) & 0 \\
       0 & 0 & D_3(E) 
     \end{array}\right], \\
   \text{where}\quad & \\
   D_i(E) &= \frac{c(2\kappa_T+R_i)}{6\kappa_T^2+3\kappa_T R_i+R_i^2},
\end{split}
\end{equation}
with $R_i = |\partial_i E|/E,\, i=1,2,3$.
We note that this function has been reformulated from its original
version \cite{HayesNorman2003} to allow increased numerical precision
for scattering-free simulations involving extremely small opacities
(i.e.~$\kappa_T=\kappa_A=\kappa\ll 1$), as is typical in cosmology
applications.

%% file: solution.tex
\section{Solving the Coupled System}
\label{sec:solution}

\subsection{Operator-Split Hydrodynamics with Radiative Feedback}
\label{sec:operator_split}

Since typical astrophysical and cosmological simulations involve
the hydrodynamic motion of gases encountering shocks, whereas radiation
diffusion and chemical kinetics processes are more of
reaction-diffusion type, we choose to solve the coupled system
\eqref{eq:cons_mass}-\eqref{eq:cons_radiation} in an operator-split
fashion.  In this approach, a time step $t^n$ to $t^{n+1}$ is taken
using the general steps
\begin{itemize}
\item[(i)] Deposit the dark matter particles onto the mesh to
  calculate the $\rho_{dm}^n$.
\item[(ii)] Solve for the gravitational potential $\phi$ resulting
  from the densities $\rhob$ and $\rho_{dm}$ using equation \eqref{eq:gravity}.
\item[(iii)] Evolve the dark matter particles using the Particle Mesh
  Method \cite{HockneyEastwood1988,NormanBryan1999,OSheaEtAl2004}.
\item[(iv)] Evolve the hydrodynamics equations
  \eqref{eq:cons_mass}-\eqref{eq:cons_energy} with a high-order,
  explicit-time upwind method. In this step, use the velocity field
  $\vb$ to advect both the chemical number densities $\mn_i$ and
  radiation energy density $E$.
\item[(v)] Using a high-order implicit-time method, solve a coupled
  reaction-diffusion system to obtain the time-evolved
  number densities $\mn_i$, radiation energy density $E$ and gas
  energy $e$.
\end{itemize}
In order to allow us to split the equations
\eqref{eq:cons_mass}-\eqref{eq:cons_radiation} into the two steps (iv)
and (v) above, we consider the gas energy as consisting of two
components, $e=\eh+\ec$, where $\eh$ is the fluid energy arising from the
hydrodynamic evolution of the system, and $\ec$ is the gas energy
{\em correction} arising from the couplings with radiation and
chemistry.  Under this decomposition, the energy conservation 
equation \eqref{eq:cons_energy} may be equivalently written as
\begin{align}
  \label{eq:cons_energy_split}
  &\partial_t (\eh+\ec) + \frac1a\vb\cdot\nabla(\eh + \ec) = \\ 
  \notag
  &\qquad -\frac{2\dot{a}}{a}\left(\eh + \ec\right)
    - \frac{1}{a\rhob}\nabla\cdot\left(p\vb\right) 
    - \frac1a\vb\cdot\nabla\phi + G - \Lambda.
\end{align}
Under this splitting, the hydrodynamic solver used in step (iv) of
the operator-split algorithm solves the system of equations 
\begin{align}
  \label{eq:enzo-cons_mass}
  \partial_t \rhob + \frac1a \vb \cdot \nabla
    \rhob &= -\frac1a \rhob \nabla\cdot\vb, \\
  \label{eq:enzo-cons_momentum}
  \partial_t \vb + \frac1a\(\vb\cdot\nabla\)\vb &=
    -\frac{\dot{a}}{a}\vb - \frac{1}{a\rhob}\nabla p - \frac1a
    \nabla\phi, \\
  \label{eq:enzo-cons_energy}
  \partial_t \eh + \frac1a\vb\cdot\nabla \eh &=
    - \frac{2\dot{a}}{a}\eh
    - \frac{1}{a\rhob}\nabla\cdot\left(p\vb\right) 
    - \frac1a\vb\cdot\nabla\phi \\
  \label{eq:enzo-chemical_ionization}
  \partial_t \mn_i + \frac{1}{a}\nabla\cdot\(\mn_i\vb\) &= 0, \\
  \label{eq:enzo-cons_radiation}
  \partial_t E + \frac1a \nabla\cdot\(E \vb\) &= 0,
\end{align}
to evolve the solution at $t^n$, $(\rhob^n,\vb^n,e^n,\mn_i^n,E^n)$, to
the time-updated variables at $t^{n+1}$,
$(\rhob^{n+1},\vb^{n+1},\eh^{n+1})$, and the advected variables
$(\mn_i^*,E^*)$.  For this step, we employ the 
Piecewise Parabolic Method (PPM) \cite{ColellaWoodward1984},
on a regular finite-volume spatial grid, implemented in the community
astrophysics code Enzo \cite{NormanEtAl2007,OSheaEtAl2004,enzo-site}. 

The remainder of the coupled system,
\begin{align}
  \label{eq:RT-cons_energy}
  \partial_t \ec &= -\frac{2\dot{a}}{a}\ec
    + G - \Lambda, \\
  \label{eq:RT-chemical_ionization}
  \partial_t \mn_i &= \alpha_{i,j} \mn_e \mn_j - \mn_i \Gamma_{i}^{ph}, \\
  \label{eq:RT-cons_radiation}
  \partial_t E &= \nabla\cdot\(D\nabla E\) - m\frac{\dot{a}}{a}E +
    4\pi\eta - c \kappa E, 
\end{align}
is then solved using a fully implicit nonlinear solution approach to
evolve the advected variables $(0,\mn_i^*,E^*)$ to the time-evolved
quantities $(e_c^{n+1},\mn_i^{n+1},E^{n+1})$.  Here we may assume
that $\ec^n=0$ since the hydrodynamic solver uses the full energy at $t^n$ in
its evolution, i.e. $\eh^n=e^n$.  Once this step is finished, we
compute the time-evolved total energy as the sum of the
hydrodynamic portion $e_h$ and the adjustments due to
radiative feedback $e_c$, i.e.~$e^{n+1} = e_h^{n+1} + e_c^{n+1}$.  The 
treatment of the implicit radiation, chemical ionization, and gas
energy feedback system
\eqref{eq:RT-cons_energy}-\eqref{eq:RT-cons_radiation} serves as our
focus for the remainder of this section.

\subsection{Solving the Radiation, Ionization and Energy Feedback System}
\label{sec:RT_solution}

Under a method-of-lines approach, we consider a two level,
up-to-second order accurate theta-scheme for implicit integration of our system
\eqref{eq:RT-cons_energy}-\eqref{eq:RT-cons_radiation},   
\begin{align}
  \label{eq:energy_simplified} 
  \ec^{n+1} + \dt\theta \mLe^{n+1} &= \ec^n + \dt(\theta-1) \mLe^{n}, \\
  \label{eq:chemistry_simplified}
  \mn_i^{n+1} + \dt\theta \mLn^{n+1} &= \mn_i^{n} + \dt(\theta-1) \mLn^{n}, \\
  \label{eq:radiation_simplified}
  E^{n+1} + \dt\theta\left[\mDE^{n+1}+\mLE^{n+1}\right] &= 
    E^{n} + \dt(\theta-1)\left[\mDE^{n}+\mLE^{n}\right].
\end{align}
Here, the parameter $\theta$ defines the implicit integration method: 
$\theta$=1 corresponds to a first-order implicit
Euler method, $\theta$=0.5 gives a second-order time-centered approach 
(i.e.~Crank-Nicolson).  We note that in the ensuing computational
results from section \ref{sec:results}, we have typically taken
$\theta$=0.51 to provide a nearly-second-order time integration while
avoiding the ``ringing'' traditionally associated with fully
time-centered approaches \cite{SimoArmero1994,Osterby2003}.
For the above equations, we have defined the diffusive operator
\begin{align}
\label{eq:mDE}
  \mDE = \mDE(E,\mn_i) &\equiv -\nabla\cdot\(D\nabla E\),
\end{align}
and we have defined the local ``reaction'' operators as
\begin{align}
  \label{eq:mLe} 
    \mLe = \mLe\(\ec,E,\mn_i\) &\equiv \frac{2\dot{a}}{a}\ec - G + \Lambda \\
  \label{eq:mLn}
    \mLn = \mLn\(\mn_i,\ec,E\) &\equiv \mn_i\Gamma_{i}^{ph} 
      - \alpha_{i,j} \mn_e \mn_j \\
  \label{eq:mLE} 
    \mLE = \mLE\(E,\ec,\mn_i\) &\equiv m\frac{\dot{a}}{a} E - 4\pi\eta + c k E.
\end{align}

The equations \eqref{eq:energy_simplified},
\eqref{eq:chemistry_simplified} and \eqref{eq:radiation_simplified}
form a coupled nonlinear system of reaction-diffusion equations for 
evolution of the fluid energy correction $\ec$, the elemental number
densities $\mn_i$, and the radiation energy density $E$.  Denoting the
vector of unknowns $U=(\ec,\mn_i,E)^T$, we first define the nonlinear
residual function for the time step $t^n \rightarrow t^{n+1}$, as
\begin{equation}
\label{eq:nonlinear_residual}
  f(U) \equiv U + \dt\,\theta
  \left(\begin{array}{c}
    \mLe(U) \\ 
    \mLn(U) \\
    \mDE(U) + \mLE(U)
  \end{array}\right)
  - \left(\begin{array}{c}
    g_{\ec}^n \\ 
    g_{\mn_i}^n \\
    g_{E}^n
  \end{array}\right),
\end{equation}
where the vectors $g_*^n$ are formed using the previous time-level
information from
\eqref{eq:energy_simplified}-\eqref{eq:radiation_simplified}.  In
order to evolve the coupled implicit system, we solve the nonlinear
problem $f(U)=0$ for the updated vector of unknowns $U^{n+1}$.  For
this nonlinear solve, we use a {\em globalized Inexact Newton's Method}  
\cite{Kelley1995,KnollKeyes2004}, in which we apply an
iterative process for convergence toward the solution $U^{n+1}$ in the
following manner.

\begin{quote}
  Given an initial guess $U_0\approx U(t^{n+1})$, we iterate
  toward a solution $U^{n+1}$ satisfying $\|f(U^{n+1})\| <
  \varepsilon \ll 1$ (we typically choose $\varepsilon=10^{-7}$):
  \begin{enumerate}
  \item Approximately solve the linearized Newton system, $\|J(U_k)
    S_k + f(U_k)\| < \delta_k$, to tolerance $\delta_k$ for the
    correction vector $S_k$.  Here, $J(U_k) \equiv
    \frac{\partial}{\partial U} f(U_k)$, and we typically choose the
    tolerance as $\delta_k = 10^{-6}\|f(U_k)\|$.
  \item Update the vector of unknowns as $U_{k+1} = U_k + \lambda_k
    S_k$, where $\lambda_k \in (\lambda_{min},1]$ is the {\em line-search
    parameter} \cite{DennisSchnabel1996,OrtegaRheinboldt2000}.
  \end{enumerate}
\end{quote}

We measure convergence of the Newton iteration with the RMS norm
\begin{equation}
\label{eq:rmsnorm}
  \|v\| = \left(\frac{\|v\|_2^2}{N(N_s+2)}\right)^{1/2},
\end{equation}
where $N(N_s+2)$ is the number of unknowns in $v$ ($N$ spatial cells,
$N_s+2$ variables), since such a norm does not grow artificially
larger with mesh refinement.
The key to efficiency of the inexact Newton algorithm lies in a fast
and robust solver for the linear systems $JS=-f$.  Once such a solver
has been provided, the algorithm exhibits very fast convergence --
superlinear for this choice of $\delta_k$
\cite{Kelley1995,EisenstatWalker1996}. Moreover, for diffusive PDE 
systems similar to the one solved here, the Newton convergence rate
has been shown to be {\em independent} of spatial resolution 
\cite{WeiserEtAl2005}, suggesting that this entire implicit
algorithm should allow scalability to the limits of the inner linear
solver.

\subsubsection{Linear Solver}
\label{subsec:lin_solver}

In solving the system \eqref{eq:nonlinear_residual}, we make one
approximation within the Newton system matrices $J(U)$, wherein we
lag the $\mn_i$ dependence of $\mDE$ in \eqref{eq:mDE} to the
previous Newton iterate.  Mathematically, this results in a full
Newton step for all but the limiter's dependence on the chemical
opacities, which are instead converged through a fixed-point
iteration.  The resulting solution retains the accuracy and stability
of the full Newton iteration, albeit with theoretically slower
convergence.  However, in practice we have not noticed any increase in
nonlinear iterations due to this approximation, and most importantly
it results in inexact Newton matrices with the form
\begin{equation}
\label{eq:jacobian_matrix}
  J(U) = I + \dt\,\theta \left[\begin{array}{ccc} 
    J_{e,e} & J_{e,\mn} & J_{e,E} \\
    J_{\mn,e} & J_{\mn,\mn} & J_{\mn,E} \\
    J_{E,e} & J_{E,\mn} & J_{E,E} \\
  \end{array}\right],
\end{equation}
where nearly all of the blocks are given by the spatially-local
components,
\begin{equation}
\label{eq:jacobian_submatrices}
\begin{array}{rclrclrcl}
  J_{e,e} &\!\!\!\equiv\!\!\!& \left[\partial_e\mLe\right] \;\;&
  J_{e,\mn} &\!\!\!\equiv\!\!\!& \left[\partial_{\mn_1}\mLe \;\; 
    \partial_{\mn_2}\mLe \;\;
    \ldots\right] \;\;&
  J_{e,E} &\!\!\!\equiv\!\!\!& \left[\partial_{E}\mLe\right] \\ \\
  J_{\mn,e} &\!\!\!\equiv\!\!\!& \left[\begin{array}{c} 
      \partial_e\mL_{\mn_1} \\ \partial_e\mL_{\mn_2} \\ \vdots
      \end{array}\right] &
  J_{\mn,\mn} &\!\!\!\equiv\!\!\!& 
    \left[\begin{array}{ccc}
      \partial_{\mn_1}\mL_{\mn_1} &
      \partial_{\mn_2}\mL_{\mn_1} & \ldots \\
      \partial_{\mn_1}\mL_{\mn_2} &
      \partial_{\mn_2}\mL_{\mn_2} & \ldots \\
      \vdots & \vdots & \ddots
    \end{array}\right] \;\;&
  J_{\mn,E} &\!\!\!\equiv\!\!\!& \left[\begin{array}{c}
      \partial_{E}\mL_{\mn_1}  \\
      \partial_{E}\mL_{\mn_2}  \\
      \vdots \end{array}\right] \\ \\
  J_{E,e} &\!\!\!\equiv\!\!\!& \left[\partial_e\mL_{E}\right] &
  J_{E,\mn} &\!\!\!\equiv\!\!\!& 
    \left[\partial_{\mn_1}\mL_{E} \;\; 
      \partial_{\mn_2}\mL_{E} \;\;
      \ldots \right],
\end{array}
\end{equation}
and the only block containing spatial couplings is 
$J_{E,E} \equiv \left[\partial_{E}\(\mD_{E}+\mL_{E}\) \right]$. 
Thus, although the Jacobian matrix contains couplings both within and
between variables, it has a very desirable structure: all
inter-variable couplings occur locally in space, and the only nonlocal
couplings are within the block $J_{E,E}$, consisting of a
scalar-valued reaction-diffusion operator.

In keeping with a block-structured view of the Jacobian
\eqref{eq:jacobian_matrix}, we rewrite the Newton system 
$JS=-f$ in the form 
\begin{equation}
\label{eq:newton_system2}
  \left[\begin{array}{cc} M & U \\ L & D \end{array}\right] 
  \left(\begin{array}{c} s_M \\ s_E \end{array}\right) = 
  -\left(\begin{array}{c} f_M \\ f_E \end{array}\right),
\end{equation}
where 
\begin{equation}
\label{eq:schur_components}
\begin{split}
  M = I + \Delta t \theta \left[\begin{array}{cc} 
      J_{e,e} & J_{e,\mn}  \\
      J_{\mn,e} & J_{\mn,\mn}
    \end{array}\right], &\qquad 
  U = \Delta t\theta \left[\begin{array}{c} J_{e,E} \\ J_{\mn,E} 
    \end{array}\right], \\
  L = \Delta t\theta \left[\begin{array}{cc} J_{E,e} & J_{E,\mn} 
    \end{array}\right], &\qquad 
  D = I + \Delta t \theta \left[\begin{array}{c} J_{E,E} \end{array}\right],
\end{split}
\end{equation}
$s_M = [s_e, s_{\mn}]^T$ and $f_M = [f_e, f_{\mn}]^T$.  We note that
the only matrix 
containing spatial dependencies is $D$, so under an appropriate
variable ordering the other sub-matrices are block diagonal.  Hence, we may
efficiently invert $M$ to obtain $s_M$ as a function of
$s_E$: 
\[
   M s_M + U s_E = -f_M \quad\Rightarrow\quad s_M = -M^{-1}(f_M + U s_E).
\]
Inserting this into the second row, we have the single equation for $s_E$,
\[
   (D - LM^{-1}U)s_E = -f_E + LM^{-1}f_M.
\]
Therefore, this {\em Schur complement} formulation
\cite{BrownWoodward2001} for solution of the linear Newton system
\eqref{eq:newton_system2} proceeds with the following steps: 
\begin{enumerate}[(i)]
\item Set $\tilde{f}_M = M^{-1} f_M$.
\item Solve the system $(D - LM^{-1}U)s_E = -f_E + L\tilde{f}_M$
  for $s_E$. 
\item Recover the remaining solution pieces, 
  $s_M = -\tilde{f}_M - M^{-1}Us_E$.
\end{enumerate}
We examine each of these steps below. 


The step (i) corresponds to solving the linear system
\begin{equation}
\label{eq:step_ia}
   \left[\begin{array}{cc} 
      I + \Delta t \theta J_{e,e} & \Delta t \theta J_{e,\mn}  \\
      \Delta t \theta J_{\mn,e} & I + \Delta t \theta J_{\mn,\mn}
    \end{array}\right]
  \left(\begin{array}{c} \tilde{f}_e \\
    \tilde{f}_{\mn} \end{array}\right) =
  \left(\begin{array}{c} f_e \\ f_{\mn} \end{array}\right).
\end{equation}
Due to the spatial locality of each component in $M$, we order
the equations and unknowns in this system so that application of $M$,
and more notably $M^{-1}$, may be performed independently in every
spatial cell.  Such solves consist of dense matrix algebra on
$(1+N_s)\times(1+N_s)$ linear systems (for $N_s$ chemical densities).
In addition, at this step we compute the matrix $M^{-1}U$ through one
additional solve with $M$, which will be used in the following steps.

The step (ii) corresponds to solving the system 
$(D - LM^{-1}U)s_E = L\tilde{f}_M - f_E$.  This is denoted the 
Schur complement system, with the matrix $\mS=D - LM^{-1}U$.  We note
that due to the spatially-local nature of the matrices $L$ and
$M^{-1}U$, we may form $\mS$ by constructing the diffusive 
sub-matrix $D$, followed by updates to the diagonal entries
corresponding to the entries of $LM^{-1}U$. Similarly, construction of
the right-hand side $L\tilde{f}_M - f_E$ may occur independently at
each spatial location. Once this system has been computed, we use a
multigrid-preconditioned conjugate gradient parallel linear solver
from the HYPRE library \cite{FalgoutYang2002,hypre-site} to perform
the scalar-valued solve, $s_E = \mS^{-1}(L\tilde{f}_M - f_E)$.  We
note that this is the only step in the solution of the Jacobian
systems \eqref{eq:newton_system2} that requires communication between 
processors.  Moreover, we point out that in recent tests the HYPRE
library has demonstrated ideal weak scaling up to over 100,000 processors for
diffusion problems similar to the one encountered in this work
\cite{BakerFalgoutYang2006}.  As this solver comprises the majority of
the non-local components within our nonlinear solver, we therefore
expect similar scalability for the overall implicit solution approach
described here.

The final step (iii) in solution of the system
\eqref{eq:newton_system2} is to recover the solution components $s_M =
(s_e, s_{\mn})^T$ via the system $s_M = -\tilde{f}_M - M^{-1} U s_E$.
Again, since we have already computed the spatially-local matrix
$M^{-1}U$ and the vector $\tilde{f}_M=M^{-1}f_M$ in step (i), we may
trivially obtain the remaining solution components through cell-local
matrix-vector products and vector operations, $s_M = -\tilde{f}_M -
(M^{-1}U)s_E$.



\subsubsection{Multiphysics/Cosmology Units}
\label{subsec:multiphysics_units}

As with any multi-physics system, special care must be taken when
solving such systems computationally due to disparate scales between
variables and equations.  This problem is especially evident in
cosmology applications, where in CGS units one may typically enounter
specific gas energies on the order of $10^{12}$, number densities on the
order of $10^{-27}$, and radiation energy densities on the order of
$10^{-15}$, with all proper density values decreasing in time due to
cosmological expansion.  To this end, we define the scaled variables
\begin{align}
  \label{eq:scalings}
  \tilde{e}_c &= \ec / u_e, \qquad \tilde{E} = E / u_E, \qquad
  \tilde{\mn}_i = \mn_i / u_n, \\
  \notag
  \tilde{x} &= x / u_x, \qquad \tilde{t} = t / u_t,
\end{align}
where the constants $u_e$, $u_E$, $u_n$, $u_x$ and $u_t$ correspond to
the typical magnitudes of gas energy, radiation energy density,
chemical number density, length and time at the start of the
simulation.  We note that due to our use of {\em comoving} values for
$E$, $\mn_i$ and $x$, these constants are all redshift-independent,
with the {\em proper} values of these quantities given by
\begin{align}
  \notag
  E_{\text{proper}} &= E / a^3(t) = \tilde{E} \frac{u_E}{a^3(t)}, \\
  \label{eq:proper_vars}
  \mn_{i,{\text{proper}}} &= \mn_i / a^3(t) = \tilde{\mn}_i \frac{u_n}{a^3(t)}, \\
  \notag
  x_{\text{proper}} &= x\, a(t) = \tilde{x}\, u_x\, a(t).
\end{align}
The constants are supplied on a problem-dependent basis, to allow for
adaptable, on-the-fly rescaling of simulations ranging from normalized
test problems to cosmological reionization.  With these rescaled
variables, we rewrite our equations
\eqref{eq:energy_simplified}-\eqref{eq:radiation_simplified} as the
normalized system
\begin{align}
  \label{eq:energy_rescaled} 
  \tilde{e}_c^{n+1} + \Delta \tilde{t}\theta \tilde{\mL}_e^{n+1} &=
    \tilde{e}_c^n + \Delta \tilde{t}(\theta-1) \tilde{\mL}_e^{n}, \\
  \label{eq:chemistry_rescaled}
  \tilde{\mn}_i^{n+1} + \Delta \tilde{t}\theta \tilde{\mL}_{\mn}^{n+1} &=
    \tilde{\mn}_i^{n} + \Delta \tilde{t}(\theta-1) \tilde{\mL}_{\mn}^{n}, \\
  \label{eq:radiation_rescaled}
  \tilde{E}^{n+1} + \Delta \tilde{t} \theta
    \left[\tilde{\mD}_E^{n+1}+\tilde{\mL}_E^{n+1}\right] &=   
    \tilde{E}^{n} + \Delta \tilde{t}(\theta-1)
    \left[\tilde{\mD}_E^{n}+\tilde{\mL}_E^{n}\right].
\end{align}
Here the operators $\tilde{\mL}_e$, $\tilde{\mL}_{\mn}$, $\tilde{\mL}_E$ and
$\tilde{\mD}_E$ have correspondingly absorbed the renormalization
constants $u_*$.  These equations, along with the normalized solution
vector $\tilde{U}=(\tilde{e}_c,\tilde{\mn}_i,\tilde{E})^T$ are then
used within the solution strategy described in section \ref{sec:RT_solution}.

\subsubsection{Adaptive Time Step Selection}
\label{subsec:adaptive_time_stepping}

A strong appeal of using implicit methods is their stability with
respect to time step size; however such freedom gives rise to the
question
of what time step should be used.
At one extreme, we may choose a large step to achieve
overall efficiency of the simulation, with little to no knowledge of
the resulting temporal accuracy.  At the other extreme, we may choose
a very small time step for an accurate solution, resulting in
inefficient simulations due to the increased cost of solving the
nonlinear systems at each step.  As the approach described here is
operator-split, in which the hydrodynamics is solved using an explicit
approach, we are therefore bound by the hydrodynamic CFL stability
limit; however for most problems involving radiation and chemical
ionization, the dynamic time scales of interest remain significantly
faster than the hydrodynamic time scale.  Thus the question of how to
adaptively choose the time step size remains.

To that end, we adaptively choose the time steps as the largest
possible that additionally satisfy a prescribed accuracy
requirement.  We estimate this accuracy through comparison of the
updated solution $U^{n+1}$ with an explicit predictor for that
solution $U^{pred}$.  Defining the weighting vector in a spatial cell
$i$ for the variable $v$ as 
\begin{equation}
\label{eq:typical_values}
   \omega_{i,v} = \sqrt{|U^{n+1}_{i,v} U^{pred}_{i,v}|} + 1,
\end{equation}
(which assumes normalized values of $U_v$), we estimate the local
accuracy of the current time step as 
\begin{equation}
\label{eq:time_error_estimator}
   \varepsilon_{loc} = \left(\frac{1}{N(N_s+2)}\left\|\frac{U^{n+1}-U^{pred}}{\omega}\right\|^p_p\right)^{1/p},
\end{equation}
where we have used the standard $p$-norm (including $p=\infty$ as the
`max' norm, in which case we do not divide by $N(N_s+2)$), and where
the quotient inside the norm is taken pointwise.
With this estimate, we set the new time step to 
\begin{equation}
\label{eq:time_step_selection}
   \Delta t^{n+1} = \frac{\tau_{\text{tol}}\,\Delta t^n}{\varepsilon_{loc}},
\end{equation}
which should provide the maximal value that still satisfies the
desired integration accuracy tolerance, $\tau_{\text{tol}}$, assuming
that $U^{pred}$ approximates the time-evolved solution $U^{n+1}$ to
$O(\Delta t)$. 

Here, the vector $\omega$ is designed so that $\varepsilon_{loc}$
estimates the average relative change in each solution component, and
includes the harmonic mean of the predicted and new states to allow
increased robustness in the case of cosmology-type problems where
variables change by orders of magnitude across cells and time steps.
The value of $p$ is typically taken to be $\infty$ in the ensuing test
problems; however such a choice may limit parallel scaling since such
a measure is sensitive to pointwise changes, of which there are many
more as dynamics propagate throughout an increasingly refined domain.  
Lastly, we use the explicit predictor as the initial guess for the
Newton method, $U_0$, which we describe in the following section
\ref{subsec:initial_guess}.

\subsubsection{Explicit Predictor}
\label{subsec:initial_guess}

A well known property of Newton's method is that its robustness and
efficiency benefit greatly from an accurate initial guess.  To this
end, we provide the predicted initial Newton iterate
\begin{equation}
\label{eq:explicit_predictor}
   U_0 = 
   \left(\begin{array}{c} e_{c,0} \\ \mn_{i,0} \\ E_0 \end{array}\right) = 
   \left(\begin{array}{c} e_c^n \\ \mn_i^n \\ E^n \end{array}\right) + \Delta t
   \left(\begin{array}{c} \mLe^n \\ \mLn^n \\ \mDE^n + \mLE^n \end{array}\right),
\end{equation}
i.e.~we use an initial guess given by the $O(\Delta t)$-accurate
explicit Euler update to the coupled system
\eqref{eq:RT-cons_energy}-\eqref{eq:RT-cons_radiation}.  As this
provides only an initial guess to the solution, its instability at
larger $\Delta t$ will not affect the
temporal stability of the overall method, since the solution to
each step must satisfy the implicit system
\eqref{eq:energy_simplified}-\eqref{eq:radiation_simplified}. 
However, as we use an adaptive time-stepping strategy, for very fast
dynamics (that give rise to very small $\Delta t$), such an initial
guess may already satisfy the nonlinear tolerance $\|f(U_0)\| <
\varepsilon$ and the solver will not require any Newton iterations,
effectively allowing an adaptive explicit/implicit simulation of the
coupled system \eqref{eq:RT-cons_energy}-\eqref{eq:RT-cons_radiation}.

\subsubsection{Adaptive Computation with Supplied Radiation Spectrum $\chi_E(\nu)$}
\label{subsec:assumed_spectrum}


The final detail that we describe in this solution approach relates to
the choice of assumed radiation spectrum $\chi_E(\nu)$.  As
noted in section \ref{subsec:fld_details}, we may choose either a
monochromatic or an integrated ``grey'' radiation equation, based on
the choice of this assumed spectrum.  This choice affects all terms
involving the radiation energy density $\Enu$ in the general radiation
energy equation \eqref{eq:cons_radiation_multifreq} and in the
coupling terms $G$ and $\Gamma_i^{ph}$.  As each of these terms
involve a product of the form $f(\nu)\Enu$, integration over $\nu$
converts these to
\begin{equation}
\label{eq:adaptive_spectrum}
   \int_{\nu_0}^{\infty} f(\nu) \Enu(\xvec,t,\nu)\, d\nu = 
   \tilde{E}(\xvec,t) \int_{\nu_0}^{\infty} f(\nu) \chi_E(\nu)\, d\nu.
\end{equation}
We therefore allow a user-defined functional form for $\chi_E(\nu)$,
which we then numerically integrate to high accuracy upon
initialization of the simulation, providing the relevant constants
necessary to convert the $\nu$-dependent equation
\eqref{eq:cons_radiation_multifreq} to the monochromatic or grey
integrated equation \eqref{eq:cons_radiation}.

%% file: results.tex
\section{Numerical Results}
\label{sec:results}

We present test problems designed to verify the accuracy of the
radiation diffusion and chemical ionization modules in conjunction 
with hydrodynamical fluid motions.  Since a number of distinct
processes may compete for importance in a full simulation, we begin
with simple tests that isolate various single components, and 
subsequently build upon those results with more sophisticated problems 
that couple additional physics.  We begin with a radiation diffusion
problem (\S\ref{sec:stream_test}) that exercises the diffusion term of 
the radiation equation \eqref{eq:cons_radiation} in the absence of
energy, chemical, or hydrodynamic coupling; this test is followed by
an examination of the matter-radiation coupling terms
(\S\ref{sec:ts_test}) in an infinite uniform medium, a diffusion wave
with material coupling (\S\ref{sec:so_test}), and a non-equilibrium
radiating shock problem (\S\ref{sec:le_test}), all of which assume
chemical equilibrium.  
Our attention then turns to problems including ionization, beginning
with an HII ionization front propagating through a static, isothermal
medium (\S\ref{sec:Iliev1_test}), followed by a cosmological I-front
propagation problem that exercises the cosmology terms and units
(\S\ref{sec:sg_test}).  We then consider a fully-coupled
radiation-hydrodynamics-ionization calculation (\S\ref{sec:wn_test}).
This section concludes with additional calculations
(\S\ref{sec:scaling_tests}) demonstrating the parallel scalability of
the radiation diffusion module, which of all components places the
highest communication demands upon a domain-decomposed parallel
calculation. 

We note that for all test problems except (\S\ref{sec:sg_test}) and
(\S\ref{sec:scaling_tests}), we use a non-cosmological problem
(i.e.~$z=0$ and $a=1$).  In problems (\S\ref{sec:sg_test}) and
(\S\ref{sec:scaling_tests}), the cosmological parameters are described
therein.

\subsection{Free-Streaming Radiation}
\label{sec:stream_test}

Because the standard diffusion equation is parabolic, the associated
signal speed of the diffusion variable is formally infinite.  However
in reality radiation fronts propagate at speeds bounded by the speed
of light in vacuum, so we modify the diffusion coefficient in our
radiation energy equation \eqref{eq:cons_radiation} with a
flux-limiter, as discussed in section~\ref{subsec:fld_details}.  Our
first test problem verifies the correct action of this limiter by
examining the propagation of a planar radiation front through a
transparent medium.  Radiation is assumed to propagate along the
$x$-axis of our computational mesh; a Dirichlet boundary condition is
imposed on the left boundary specifying an incident radiation energy
density of 1.0 erg cm$^{-3}$.  Physically, the expectation is that
with a sufficiently small (but nonzero, due to numerical constraints)
opacity, a sharp radiation front will move through the domain at the
speed of light.  The Planck and Rosseland mean opacities are assigned
a constant value of $10^{-6}$ cm$^{-1}$, ensuring an essentially
transparent medium.  The spatially uniform initial value of the
radiation energy density is 10$^{-4}$ erg cm$^{-3}$.

\begin{figure}
 \centerline{
   \includegraphics[width=0.8\linewidth]{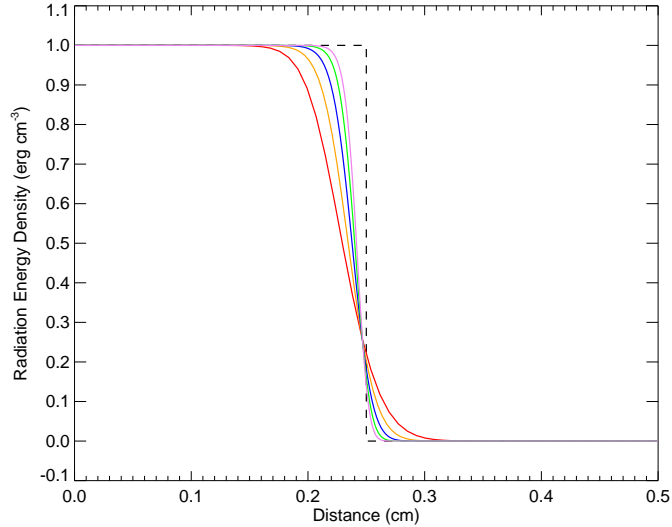}}
 \caption{Curves of $E$ vs. $x$ for mesh sizes of 128 (red), 256
   (orange), 512 (blue), 1024 (green), and 2048 (violet) zones.  The
   analytical solution (black dashed line) is a step function centered
   at $x$ = 0.25cm.}
 \label{fig:stream05}
\end{figure}
The computational mesh has a domain length of 1.0 cm along the
propagation direction of the light wave.  We have run the problem for
8.3391 picoseconds, which is one quarter of the light-crossing time
for this length. Figure~\ref{fig:stream05} shows a series of curves
resulting from calculations at mesh sizes of 128, 256, 512, 1024, and
2048 zones along the $x$-axis.  The dashed line indicates the expected
location of the radiation front, $c\, t$, where $c$ is the speed of light
and $t$ the evolution time.  In the absence of the flux limiter, the
numerical curves would give the formal $t\!\to\!\infty$
solution for the diffusion equation, which for our problem parameters
would be a nearly horizontal profile for $E$ throughout the domain
(given the nearly zero opacity).  That the curves capture the
correct location of the radiation front is due entirely to the action
of the limiter.  The sharpening of this front with increased
resolution is evident.  

Also apparent is a slight lag (about 0.01 cm) between the analytical
location of the radiation front and the numerical location, taken as
the common intersection point of the numerical curves. The size of
this lag depends upon the choice of the adaptive time step.  In the
language of (\S\ref{subsec:adaptive_time_stepping}), we compute
$\Delta t$ using $p=\infty$ and we vary $\tau_{\text{tol}}$, which
here corresponds to the maximum allowed fractional change in the
radiation energy density per timestep. 
\begin{figure}
 \centerline{
   \includegraphics[width=0.8\linewidth]{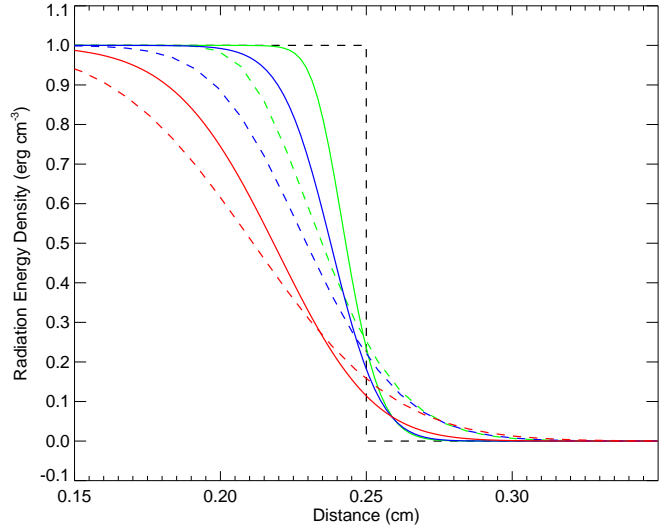}}
 \caption{Radiation energy profiles at 128 (dashed curves) and 512
   (solid curves) zones for $\tau_{\text{tol}}$ of 0.01 (green), 0.05
   (blue), and 0.25 (red).  Intersections of associated 128 and
   512-zone curves indicate the ``consensus'' location of the
   radiation front for a given choice of $\tau_{\text{tol}}$.
   Convergence to the correct value is readily observed.}
 \label{fig:streamlag}
\end{figure}
Figure~\ref{fig:streamlag} illustrates this effect by showing 3 curve
pairs, each of which has been computed with a different choice of
$\tau_{\text{tol}}$: 0.01 (green), 0.05 (blue), and 0.25 (red).  In
each colored pair, the solid curve was obtained using 512 zones, and
the dashed curve shows the 128-zone result.  Two curves at different 
mesh resolutions are provided to identify the ``consensus'' value of
the light front location via their point of intersection.  This
location is seen to converge as $\tau_{\text{tol}}\to0$.

\subsection{Matter-Radiation Equilibration in a Homogeneous Medium}
\label{sec:ts_test}

We now consider a problem in which $e$ and $E$ are spatially uniform
but are initialized to values far away from equilibrium.  This case
thus isolates the matter-radiation coupling terms in the gas and
radiation energy equations.  The parameters for this test were
published by Turner and Stone~\cite{TurnerStone2001}, who assumed an
isotropic medium characterized by a single opacity of $4 \times
10^{-8}$ cm$^{-1}$, a gas density of 10$^{-7}$ g cm$^{-3}$, and an
average particle mass of 0.6 m$_H$, where m$_H$ is the mass of a
hydrogen atom. Coupled
to this medium is a radiation field with a uniform value of 10$^{12}$
erg cm$^{-3}$.  From this value we compute a ``radiation
temperature'', $T_r \equiv (E/a_r)^{1/4}$, of about $3.4 \times 10^6$
K.  Here we have defined $a_r$ as the radiation constant, $7.56 \times
10^{-15}$ erg cm$^{-3}$ K$^{-1}$.  Two cases are considered: one in
which the initial gas energy density is $e = 10^{10}$ erg cm$^{-3}$, and
one in which the initial value is $e = 10^2$ erg cm$^{-3}$.  For the
stated parameters, these energies correspond to gas temperatures of
roughly $4.8 \times 10^8$ and 4.8 K, respectively, which therefore
bracket the radiation temperature.  In both cases, however, the
initial radiation temperature is sufficiently high that the radiation
energy density should remain constant to good approximation as the gas
evolves to thermal equilibrium.  To see this clearly, consider the
effective heat capacity of a unit volume of the radiation ``gas'' as
compared to that for the material.  For radiation, this number is
simply 1.0 cm $\times$ $a_r T_r^3$, which evaluates to roughly $3
\times 10^5$ erg K$^{-1}$.  In contrast, the gamma value and mean
particle mass translate to a {\it specific} heat (erg g$^{-1}$
K$^{-1}$) of roughly $2.0 \times 10^8$, which yields, for our assumed
density, a material heat capacity of 20 erg K$^{-1}$ for a unit
control volume.  The physical result is that the radiation field has
an effectively infinite thermal reservoir when compared to the
material. 

\begin{figure}
 \centerline{
   \includegraphics[width=0.8\linewidth]{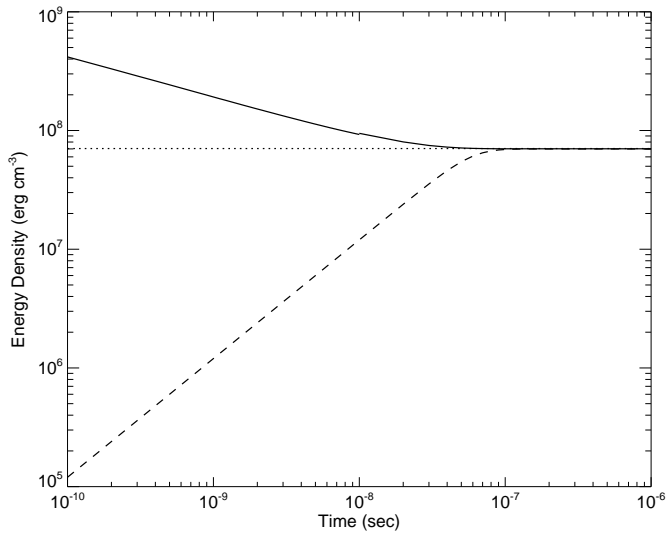}}
 \caption{Evolution to thermal equilibrium of a medium with an
   initially high (solid line) and low (dashed line) gas energy
   density.  The analytical equilibrium value is shown by
   the horizontal dotted line.}
 \label{fig:turnerstone}
\end{figure}

If the radiation energy density is formally assumed to be constant,
the gas energy equation may be written as a simple ODE:
\begin{equation}
\label{eq:TS-gas_ode}
  \dot{e} = c\kappa\,E - 4\pi\kappa\,B(e),
\end{equation}
where $B$ is the temperature-dependent Planck function
\begin{equation}
\label{eq:TS-blackbody}
  B(T) = \frac{c\,a_r}{4\pi}\,T^4.
\end{equation}
Using the ideal-gas law \eqref{eq:eos}, we write $B$ as a function of
$e$ and solve the simplified gas energy equation for the equilibrium
value of $e$ such that $\dot{e} \equiv 0$, 
\begin{equation}
\label{eq:TS-equilibrium}
  e_{\rm eq} = \frac{3}{2}\left(\frac{\rho\,k_B}{0.6\,m_H}\right) 
    \left(\frac{E}{a_r}\right)^{1/4}.
\end{equation}
Notice that this expression is nothing more than the ideal gas formula
for $e$ evaluated at the fixed radiation temperature, the expected
result.  

The results of our two test calculations are shown in
figure~\ref{fig:turnerstone}. Both tests were run in a small box
domain (4$^3$ zones) with triply-periodic boundaries. The case of
$T(0) > T_r$ is indicated by the solid curve; the low-$T$ case is
shown by the dashed line.  The horizontal dotted line has been placed
at $e = e_{\rm eq}$.  Both energy curves converge to the correct
result.  Note that in this test the opacity serves only to control the
timescale to reach thermal equilibrium; neither the value of the
equilibrium energy nor the validity of our assumption of constant $E$
are dependent on the value of $\kappa$.  
While figure~\ref{fig:turnerstone} demonstrates convergence to the
correct asymptotic value of the gas energy, it provides no information
as to whether the {\it rate} at which it approaches this value is
correct. A quantitative assessment of this latter metric is provided
by our next test problem. 

We further use this test to examine the conservation properties of the
coupled radiation and gas energy solver.  In Table
\ref{table:conservation} we show the value of 
\begin{equation}
\label{eq:cons_energy_test}
   \frac{\int \left| E_{total}(t) - E_{total}(0)\right|\,\mathrm{d}x}
   {\int E_{total}(0)\,\mathrm{d}x}
\end{equation}
for both tests at the final time $t$=2.5e-7 sec., run using a variety
of nonlinear and linear solver tolerances, $\varepsilon$ and $\delta$,
respectively.  We note that in all cases, the total energy is
conserved to more than 10 digits of accuracy.  Moreover, while the
conservation is weakly dependent on the nonlinear solver tolerance, it
is entirely independent of the linear solver tolerance.  This behavior
is most likely due to use of the Schur complement formulation
(\S\ref{subsec:lin_solver}), that exactly solves for coupling between 
variables to floating point roundoff, leaving the iterative linear
solver to handle only the radiation equation.
\begin{table}
{\scriptsize
  \centering
  \begin{tabular}{|c|c|c|c|} \hline
     $\varepsilon$ & $\delta$ & $L\to H$ error & $H\to L$ error \\
     \hline \hline
     $10^{-7}$  & $10^{-9}$  & $7.04\times10^{-12}$ & $1.37\times10^{-12}$  \\
     $10^{-10}$ & $10^{-9}$  & $3.07\times10^{-13}$ & $3.45\times10^{-14}$  \\
     $10^{-4}$  & $10^{-9}$  & $3.55\times10^{-11}$ & $1.83\times10^{-11}$  \\
     $10^{-7}$  & $10^{-12}$ & $7.04\times10^{-12}$ & $1.37\times10^{-12}$  \\
     $10^{-7}$  & $10^{-3}$  & $7.04\times10^{-12}$ & $1.37\times10^{-12}$  \\
     \hline
  \end{tabular}
  \caption{Conservation of total energy for the matter-radiation
    equilibration test.  Relative error in energy conservation
    (\ref{eq:cons_energy_test}) for both low-to-high and high-to-low
    temperature equilibration, for various nonlinear and linear solver
    tolerances $\varepsilon$ and $\delta$, respectively.}
  \label{table:conservation}
}
\end{table}
We further note that this is an ideal problem to test conservation of
the coupled solver, since it is the only test considered that uses a
closed system.  We further comment that
since the PPM finite-volume method is constructed to satisfy
conservation, and the implicit subsolver achieves conservation to high
accuracy, overall conservation of the coupled solver follows.
However, we note that for problems utilizing non-periodic boundary
conditions, chemical ionization cooling, gravitational heating, or
cosmological expansion, the model no longer represents a closed system
and therefore will not conserve energy.

\subsection{Non-Equilibrium Marshak Waves}
\label{sec:so_test}

This test exercises both radiation diffusion and the physics of
matter-radiation coupling.  Non-equilibrium Marshak waves characterize
the evolution of the radiation field in an initially cold, uniform
halfspace on which a radiation source is imposed.  The particular form
of the Marshak problem described here is originally due to
Pomraning~\cite{Pomraning1979}.  The problem was re-examined by Su and
Olson~\cite{SuOlson1996}, who derived semi-analytic exact solutions
for the radiation and gas energy densities and tabulated select values
of them on a grid of space and time values.  The problem considers a
formally 1-D semi-infinite domain in which $(z,t)$ denote dimensional
space and time coordinates.  Ignoring hydrodynamic motions, Su and
Olson write simplified forms for the radiation and material energy
equations as
\begin{align}
  \label{eq:suolson_rade}
  \partial_t E(z,t) - \partial_z\left[\frac{c}{3\kappa}\partial_z E(z,t)\right]
    &= c\kappa\left[a_r T^4(z,t) - E(z,t)\right], \\
  \label{eq:suolson_gast}
  c_v(T) \partial_t T(z,t) &= c\kappa\left[E(z,t) - a_r T^4(z,t)\right],
\end{align}
in which $\kappa$ is the constant opacity, $a_r$ the radiation
constant as defined previously, and $c_v$ the specific heat of the
material.  Note that the flux-divergence term in
\eqref{eq:suolson_rade} assumes pure diffusion with no flux
limiter.  The matter temperature, $T$, is assumed to be related to the
gas energy, $e$, via 
\begin{equation}
  \label{eq:suolson_eos}
  e \ = \ \rho c_v T.
\end{equation}
As written, \eqref{eq:suolson_rade} and \eqref{eq:suolson_gast} are
coupled nonlinear PDEs in the dependent variables $E$ and $T$.  Su
and Olson linearized the equations by choosing the following form for
the specific heat: 
\begin{equation}
  \label{eq:suolson_spheat}
  c_v \ = \ \alpha T^3,
\end{equation}
where $\alpha$ is an arbitrary constant.  The $T^3$ dependence of
$c_v$ on temperature has two effects: it allows equations
\eqref{eq:suolson_rade} and \eqref{eq:suolson_gast} to be written as
linear ODEs in $E$ and $T^4$, and it gives the heat capacity of the
material the same temperature dependence as the effective heat
capacity of the radiation field.  With an appropriate choice of
$\alpha$, a problem can therefore be designed in which the material
and radiation will both evolve significantly in space and time.

The description of the problem is completed with a specification of
the boundary conditions. A Marshak boundary condition is applied to
$E$ at $z=0$: 
\begin{equation}
  \label{eq:suolson_leftbc}
  E(0,t) \ - \ \frac{2}{3\kappa} \partial_z E(0,t) \ = \ \frac{4}{c}F_{\rm inc},
\end{equation}
where $F_{\rm inc}$ is the incident flux at $z=0$.  The boundary
condition at $z=\infty$, and initial conditions at $t=0$ are
\begin{equation}
  \label{eq:suolson_rightbc_initcond}
  E(\infty,t) \ = \ 0, \qquad 
  E(z,0) \ = \ T(z,0) \ = \ 0.
\end{equation}

\begin{figure}
 \centerline{
 \includegraphics[width=0.8\linewidth]{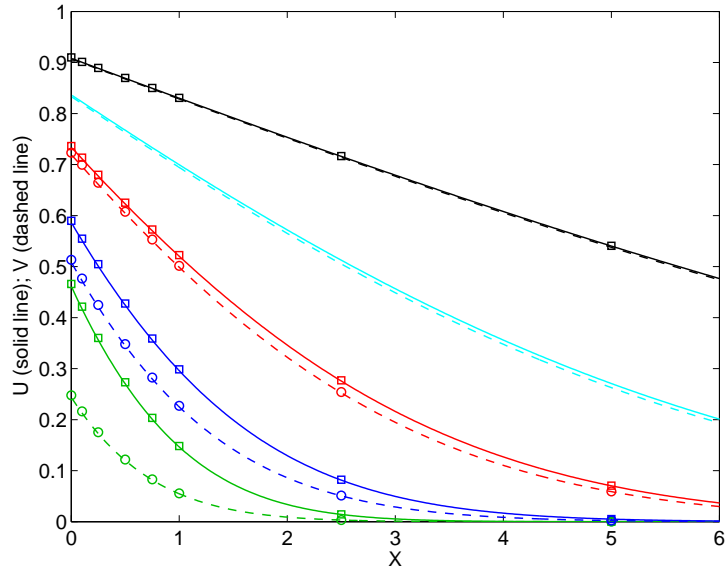}}
 \caption{Curves of $e$ (dashed line) and $E$ (solid line)
   vs.~dimensionless distance, at dimensionless times
   $\tau$ = 1, 3, 10, 30, and 100 (curves shift upward as $\tau$
   increases).  Analytic values for $e$ (squares) and $E$ (circles)
   are shown for $\tau$ = 1, 3, and 10; values for
   $E$ are also provided at $\tau=100$, by which time $e$ and $E$
   are nearly equilibrated.} 
 \label{fig:Marshak2048}
\end{figure}

Su and Olson construct the linearized equations by defining
dimensionless independent and dependent variables, $(X,\tau)$ and
$(u,v)$ such that
\begin{align}
  \label{eq:SO-dimensionless_vars}
  X \ &\equiv\ z \kappa \sqrt{3}, &
  u(X,\tau) \ &\equiv \ \left(\frac{c}{4\,F_{\rm inc}}\right) E(z,t),
  \\
  \notag
  \tau \ &\equiv \ \left(\frac{4a_r c\kappa}{\alpha}\right)t, &
  v(X,\tau) \ &\equiv \ \left(\frac{c\,a_r}{4\,F_{\rm inc}}\right)T^4(z,t).
\end{align}
With these definitions, and letting $\epsilon = 4\,a_r/\alpha$,
equations \eqref{eq:suolson_rade}, \eqref{eq:suolson_gast}, and
\eqref{eq:suolson_leftbc}-\eqref{eq:suolson_rightbc_initcond} become 
\begin{align}
  \label{eq:suolson_nondimrad}
  \epsilon\,\partial_{\tau} u(X,\tau) \ - \ 
    \partial^2_{X^2} u(X,\tau) \ &= \ v(X,\tau) \ - \ u(X,\tau), \\
  \label{eq:suolson_nondimgas}
  \partial_{\tau} v(X,\tau) \ &= \ u(X,\tau) \ - \ v(X,\tau), \\
  \label{eq:suolson_nondimleftbc}
  u(0,\tau) \ - \ \frac{2}{\sqrt{3}}\,\partial_X u(0,\tau) \ &= \ 1, \\
  \label{eq:suolson_nondimrightbc}
  u(\infty,\tau) \ &= \ 0, \\
  \label{eq:suolson_nondiminitc}
  u(X,0) \ = \ v(X,0) \ &= \ 0.
\end{align}

\begin{figure}
 \centerline{
   \includegraphics[width=0.8\linewidth]{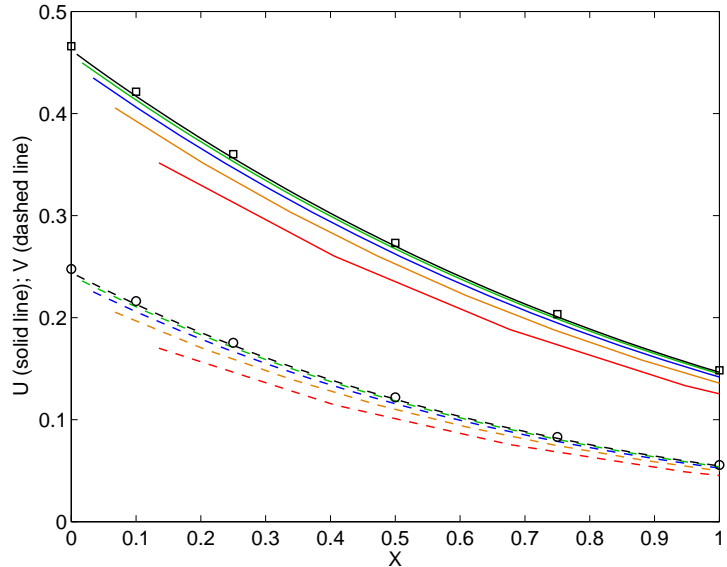}}
 \caption{Marshak problem convergence: curves of $u$ (solid lines) and
   $v$ (dashed lines) at mesh resolutions of 128 (red), 256 (orange), 512
   (blue), 1024 (green) and 2048 (black) zones.  Reference solution
   values are indicated by open circles and squares; the halving of
   relative error with each doubling of mesh size is readily apparent.} 
 \label{fig:MarshakM}
\end{figure}

The Marshak boundary condition represented by
\eqref{eq:suolson_nondimleftbc} enforces the constraint of constant
flux on the left boundary.  This is an example of a ``mixed'' or Robin
boundary condition, and as such requires special treatment in Enzo.
For the purposes of this verification test, we implement this boundary
condition by imposing a Dirichlet condition 
with a time-varying value of $u$ computed from \cite{SuOlson1996}'s
equation 36, evaluated at $X=0$.  Because the integrands in their
equation are highly oscillatory for $\tau \ll 1$, we substitute the
asymptotic expression given by their equation 51 when $\tau<10^{-5}$. 

Figure~\ref{fig:Marshak2048} shows results from a high-resolution
simulation with 2048 zones along the $X$ coordinate.  
The exact solution values tabulated by \cite{SuOlson1996} span the
range $0 \leq X \leq 10$.  Since the right boundary condition is
specified at $X = \infty$, we choose our domain $X\in[0,L]$ such that
$L$ is sufficiently large (about 35) for the evolution 
time of interest that the Dirichlet condition $X(L)=0$ may be
reasonably applied. 
We choose opacity and coupling parameters $\kappa = \epsilon =$ 1.0
cm$^{-1}$.
The curves indicate profiles of $u$ (dashed lines) and $v$ (solid
lines) for $\tau$ values of 1, 3, 10, 30, 100.  The squares and
circles indicate exact values of $u$ and $v$, respectively.  We have
indicated these values on corresponding curves at evolution times
sufficiently early that the material and radiation have not yet had
time to equilibrate.  Figure~\ref{fig:MarshakM} shows a resolution
study for the curves computed at $\tau=1$.  Curves at mesh sizes of
128 (red), 256 (orange), 512 (blue), 1024 (green), and 2048 (black)
are shown.  Each calculation is performed with the timestep
restriction $\tau_{\text{tol}} = 0.05$.  Because this treatment allows
for adaptive timesteps, the evident first-order rate of convergence
measures the combined effect of time and space discretization methods.

\subsection{Subcritical Radiating Shock Waves}
\label{sec:le_test}

We now add hydrodynamic motions to our mix of physics by examining the
propagation of shock waves for which the radiation energy plays a
significant role in the shock structure and evolution.  Radiating
shock waves represent a broad class of phenomena figuring prominently
in both astrophysical and terrestrial applications.  The particular
formulation of the problem we present is due to Lowrie and Edwards
\cite{LowrieEdwards2008}, who considered the propagation of planar,
steady shock waves in the grey nonequilibrium diffusion limit.  Under
the assumption of steady flow, \cite{LowrieEdwards2008} transform the
coupled gas and radiation energy equations into a set of nonlinear
ODEs in dimensionless gas and radiation temperature variables, which
must be integrated numerically to achieve semi-analytic solutions.
Nonetheless, their radiation diffusion model corresponds identically
to that implemented in Enzo in the grey LTE limit, and the unique
structure of the post-shock material temperature profile for a given
Mach number makes this problem an excellent verification test for
computer codes.

\begin{figure}
 \centerline{
   \includegraphics[width=0.8\linewidth]{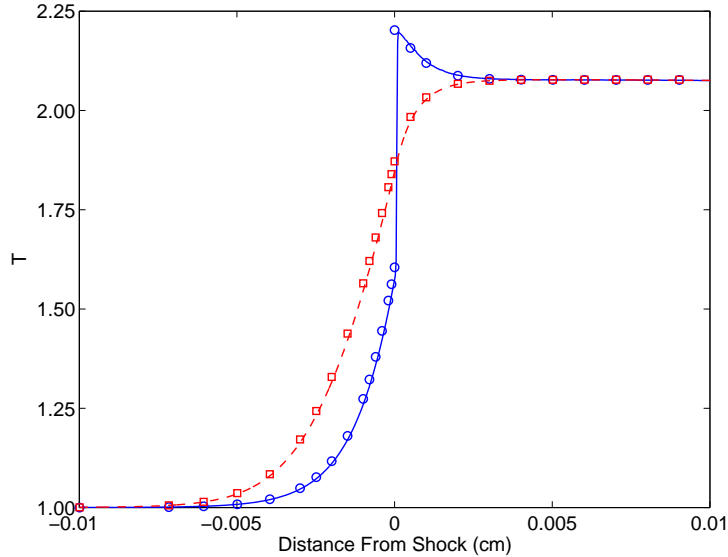}}
 \caption{Subcritical radiating shock test using 4096 spatial zones.
   Gas and radiation temperatures are plotted in units of their
   preshock values: $T$ is the solid curve; $T_r$ is the dashed curve.
   Semi-analytic values for $T$ and $T_r$ are indicated by circles and
   squares, respectively.}
 \label{fig:rshock}
\end{figure}

We have run the Mach-2 test case described in
\cite{LowrieEdwards2008}.  The computational domain has a length of
0.1 cm.  The material has a uniform initial density of 1.0 g
cm$^{-3}$, a constant specific heat of $2.218056\times 10^{12}$ erg
g$^{-1}$ eV$^{-1}$, and a uniform initial velocity of $1.9475 \times
10^5$ cm s$^{-1}$.  The material and radiation are assumed to be in
thermal equilibrium at $t=0$ at a temperature of 121.6 eV. Outflow
and reflecting boundary conditions are imposed upon the left and right
boundaries, respectively, resulting in a shock wave that forms near
the right boundary and propagates to the left.  The total evolution
time is 1.73325 nanoseconds.

\begin{figure}
 \centerline{
   \includegraphics[width=0.8\linewidth]{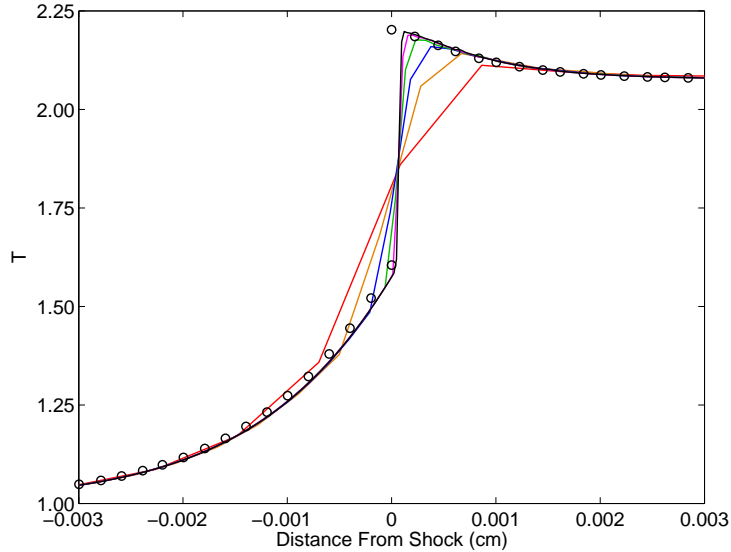}}
 \caption{Subcritical radiating shock convergence: gas temperature vs
   distance from the shock for mesh resolutions of 128 (red), 256
   (orange), 512 (blue), 1024 (green), 2048 (cyan) and 4096 zones
   (black). Semi-analytic reference values are indicated by open
   circles.}
 \label{fig:rshockRES}
\end{figure}

Figure~\ref{fig:rshock} shows the result of a high-resolution
simulation (4096 zones along the propagation axis).  The curves represent
the dimensionless gas (solid curve) and radiation (dashed curve)
temperatures, $T$ and $T_f$.  The circles and squares are taken from
exact solution data kindly provided by R. Lowrie for this parameter
set.  Both the gas and radiation have dimensionless far-field
temperatures of 1.0 in the pre-shock state.  Examining the gas
temperature curve, there are three features of particular interest:
the {\it precursor}, in which the material is preheated ahead of the
shock front by the radiation wave which travels ahead of the shock;
the {\it Zel'dovich spike}, shown by the overshoot in temperature at
the shock front, and the {\it radiation relaxation region}, delineated
by the decline in the material temperature to its eventual far-field
postshock value.  Letting $T_p$ denote the maximum preshock value of
the gas temperature in the precursor, and $T_1$ the asymptotic
postshock value, we note that the property $T_p < T_1$ identifies this
calculation as an example of a {\it subcritical}
radiating shock.  In the limit of high Mach number, $T_p$ can become
equal to (but never exceed) $T_1$, such a shock wave is referred to as
{\it supercritical}. 

As vividly demonstrated by \cite{LowrieEdwards2008}, the strength of
the precursor, the height of the Zel'dovich spike, and the precise
temperature structure in the relaxtion region are extremely sensitive
to the Mach number.  While the case we have shown is subcritical, it 
lies 
near the limit for which a multidimensional code can reasonably
capture this structure without resorting to adaptive mesh refinement.
The degree to which we resolve this structure as a function of
resolution is shown in figure~\ref{fig:rshockRES}, in which we magnify
the region near the shock and show gas temperature curves for mesh
sizes of 128, 256, 512, 1024, 2048 and 4096 zones.  As shown in
\cite{LowrieEdwards2008}, raising the Mach number results in a
dramatic increase in the height of the spike and narrowing of the
relaxation region; a proper representation of the postshock structure
in a supercritical shock with Enzo must await the implementation of
adaptive mesh refinement in our radiation module. 

Since this problem considers coupled radiation and hydrodynamics, we
also examine how the adaptive time step selection strategy from
Section \ref{subsec:adaptive_time_stepping} compares with the
hydrodynamic CFL-limited time step.  For this problem, the average
radiation time step ranged from 1.1e-5 down to 9.4e-5 for the coarsest
(128-cell) to finest (4096-cell) grids, exhibiting a near-constant
time step selection that tracks evolution of the radiation field.  For
these same problems, the hydrodynamic CFL limits on the time steps
were 3.3e-4, 1.7e-4, 4.2e-5, 2.1e-5, 1.0e-5 and 5.0e-6.  Hence, for
most problems the stiff radiation time scale limits the overall
time step size, until very fine grids where the mesh-dependent
hydrodynamic CFL stability condition becomes more restrictive.

\subsection{Isothermal Ionization of a Static Neutral Hydrogen Region}
\label{sec:Iliev1_test}

Our first test problem incorporating ionization chemistry is due to
Iliev et al.~\cite{IlievEtAl2006}.  This problem combines radiative
transfer and hydrogen ionization in a static astrophysical
region.  The physical situation of interest is the expansion of an
ionized hydrogen (HII) region in a uniform gas around a single
monochromatic ionizing source emitting
$\dot{N}_{\gamma}=5\times10^{48}$ photons per second at the ionization
frequency of hydrogen ($h\nu = 13.6$ eV).  We enforce a fixed gas
temperature of $T=10^4$ K, and a static hydrodynamic state
(i.e.~$\dot{\rhob}=\dot{\vb}=\dot{e}=0$).  In such a problem, the radiation
source should rapidly ionize the surrounding hydrogen, and then
should develop a spherically-propagating ionization front (I-front)
that propages quickly at first, slows, and then 
eventually stagnates at an equilibrium position referred to as the
Str{\"o}mgren radius, where ionization (HI$\to$HII) and recombinations
(HII$\to$HI) balance.  For this scenario, the analytically-provided
I-front radius is given by 
\begin{align}
  \label{eq:Ifront_solution}
  r_I &= r_S [1-\exp(-t/t_{rec})]^{1/3}, \quad\text{where}\\
  \label{eq:Stromgren_radius}
  r_S &= \left[\frac{3\dot{N}_{\gamma}}{4\pi \alpha_B \mn_H^2}\right]^{1/3},
\end{align}
and the recombination time is given by $t_{rec}=(\alpha_B\mn_H)^{-1}$.
Here, $\alpha_B$ = $2.59\times10^{-13}$ cm$^3$ s$^{-1}$ is the case B
hydrogen recombination coefficient.  

We have the following problem parameters: the domain size is $L=6.6$
kpc in each direction; the initial gas number density is $\mn_H =
10^{-3}$ cm$^{-3}$; the initial radiation energy density is $E =
10^{-20}$ erg cm$^{-3}$; the initial ionization fraction (HII/H) is
0.0012; the ionization source is located in the lower corner of the
box (the $(1,1,1)$ cell); we use reflecting boundary conditions at the
$x$-, $y$- and $z$-left boundaries, and outflow conditions at the
corresponding right boundaries.  For these parameters the
Str{\"o}mgren radius $r_S = 5.4$ kpc, the recombination time $t_{rec} 
\approx$ 3.86e15 s ($\approx$ 122.4 Myr), and the total simulation
time is 500 Myr ($\approx 4 t_{rec}$).  The implicit solver parameters
used were a convergence norm of $p=2$, desired solution tolerance of
$\tau_{\text{tol}}=0.01$, time-step parameter of $\theta=0.51$, and
nonlinear solver tolerance of $\varepsilon=10^{-7}$.

\begin{figure}
\centering
\parbox{\columnwidth}{
\subfigure{\includegraphics[width=0.5\columnwidth]{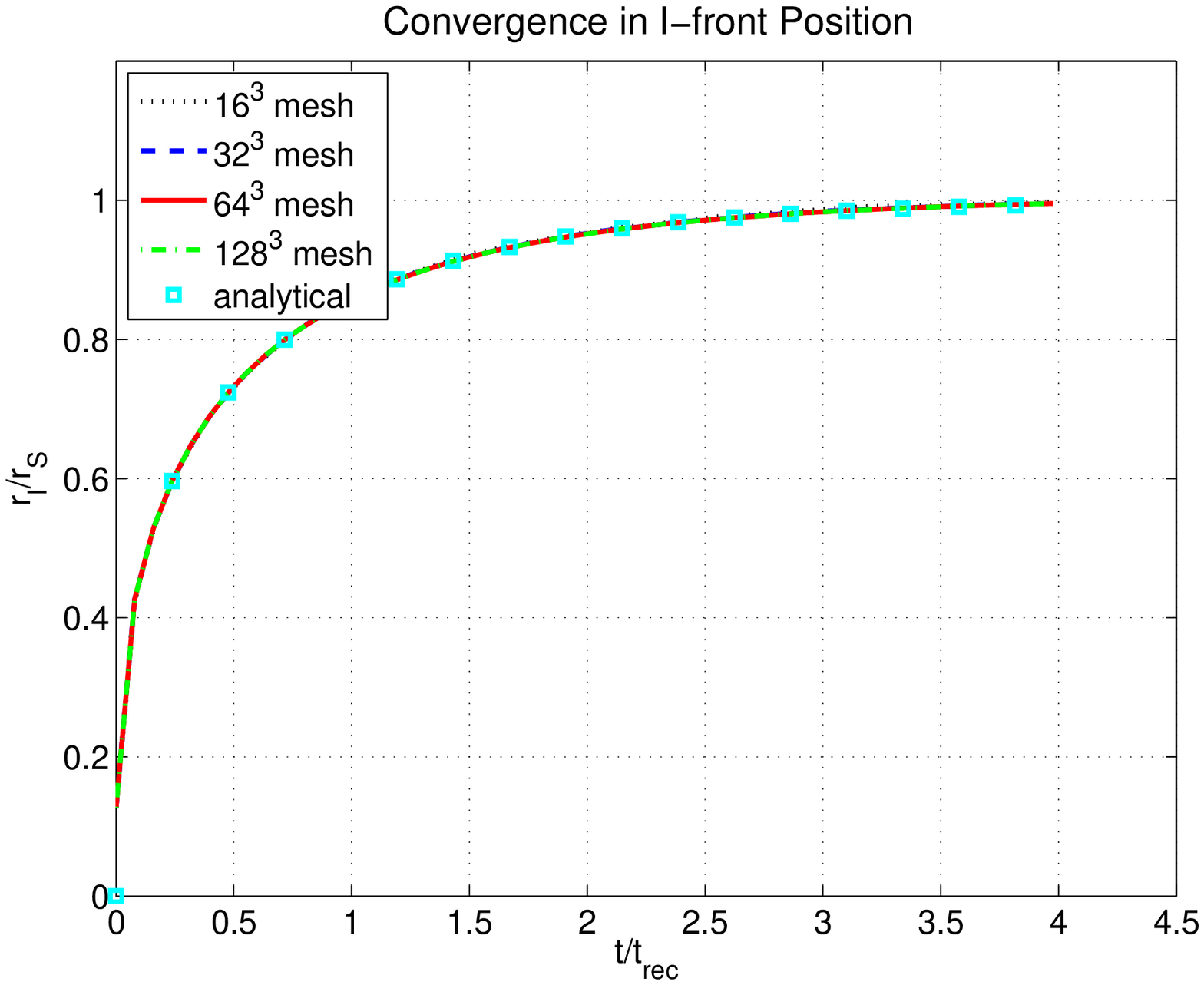}}
\hfill
\subfigure{\includegraphics[width=0.5\columnwidth]{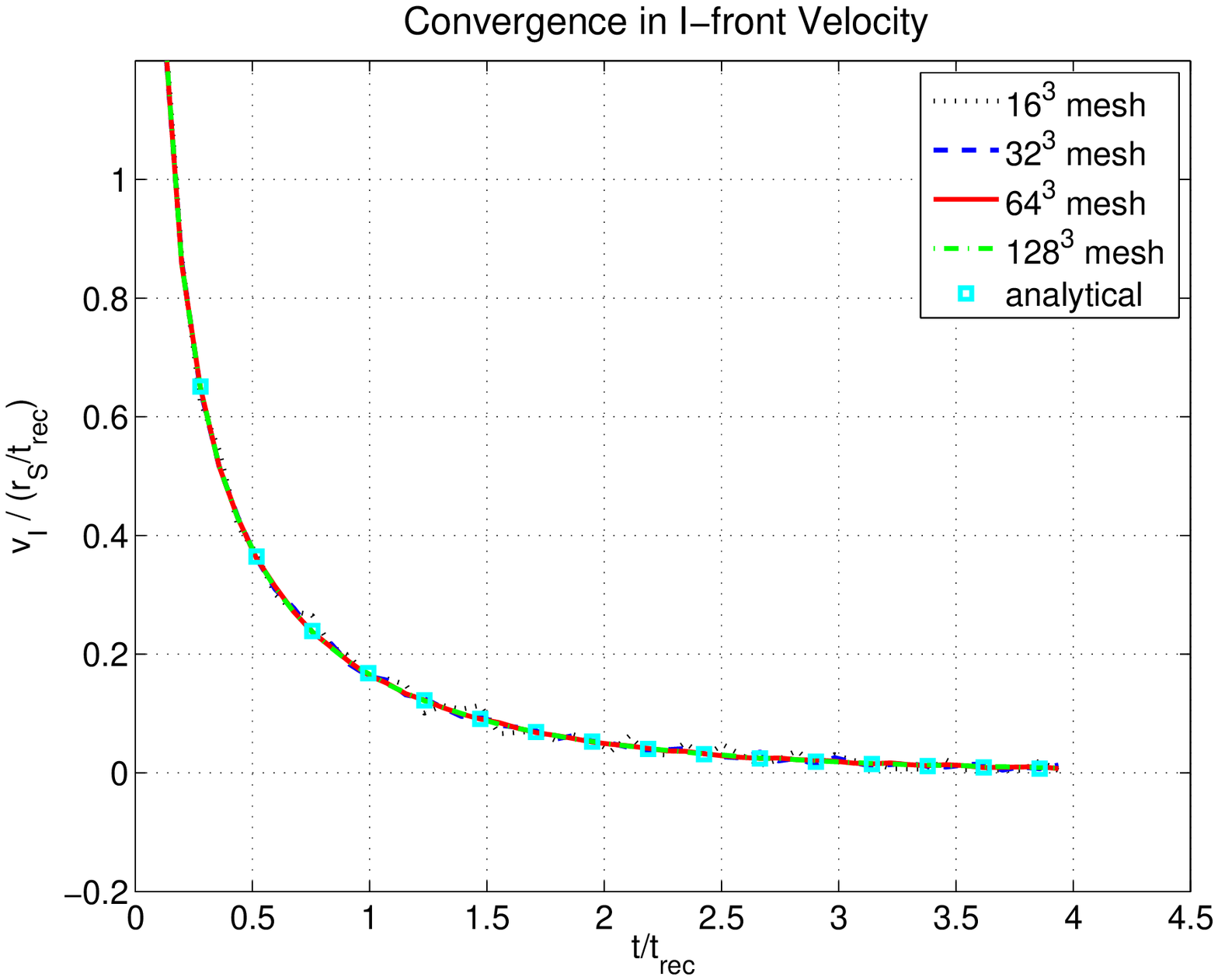}}
}
\parbox{\columnwidth}{
\subfigure{\includegraphics[width=0.5\columnwidth]{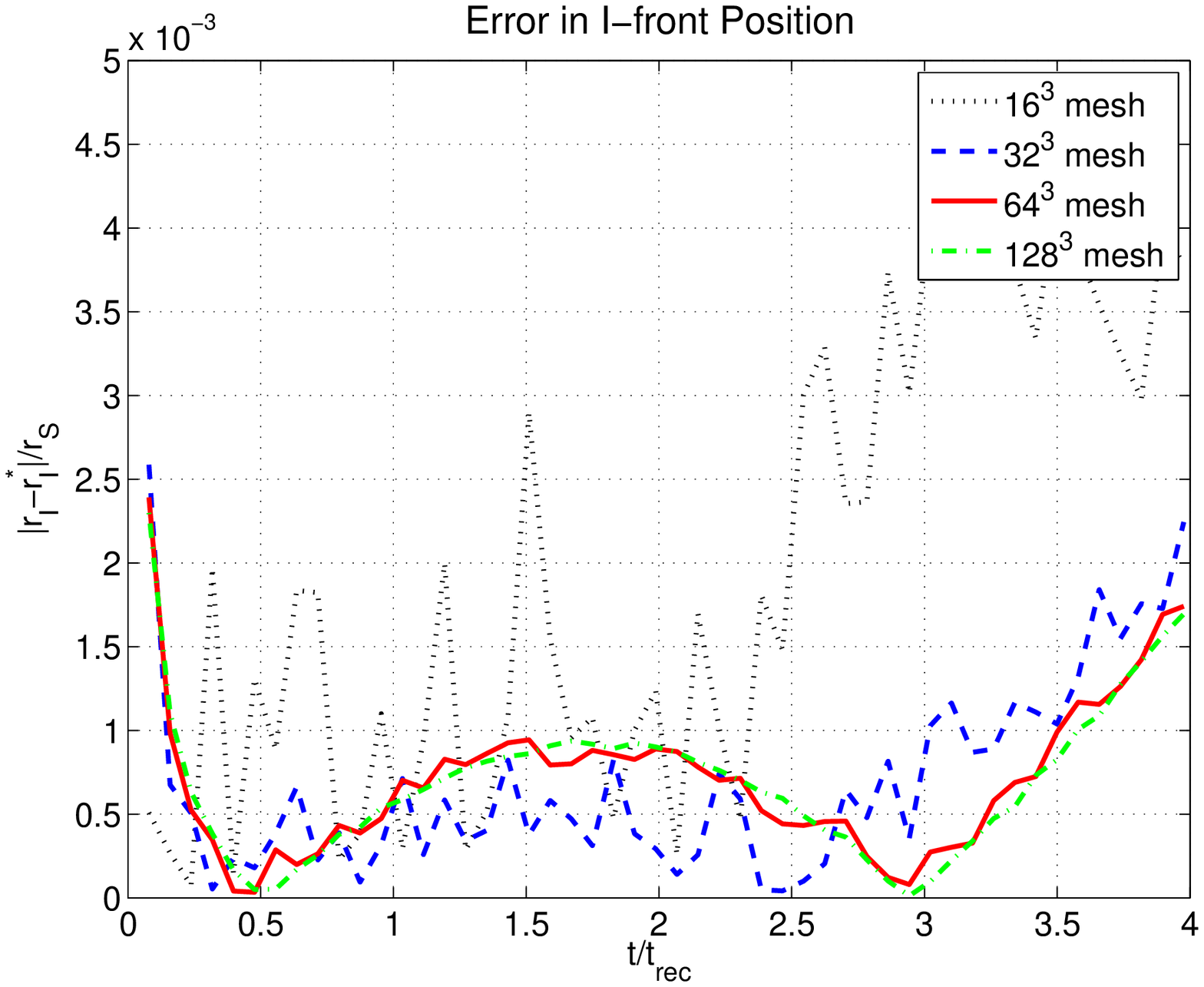}}
\hfill
\subfigure{\includegraphics[width=0.5\columnwidth]{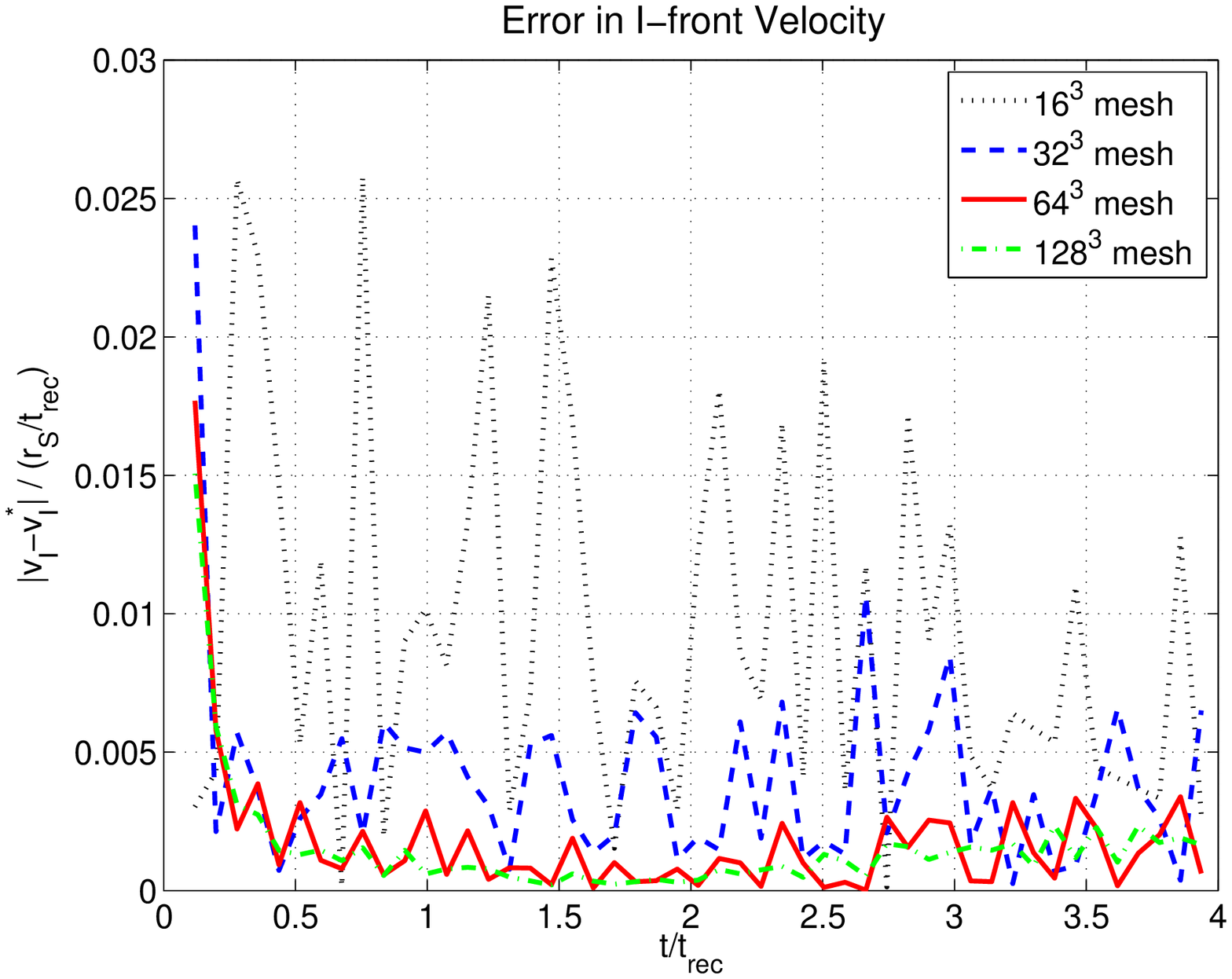}}
}
  \caption{Left: evolution of the I-front position, analytical
    solution, and relative error for varying resolutions.  Right:
    computed I-front velocity and relative error.} 
  \label{fig:IlievEtAl1_propagation}
\end{figure}

In Figure \ref{fig:IlievEtAl1_propagation} we plot the
spherically-averaged I-front position and radius with respect to time,
for various spatial mesh sizes.  The I-front position is computed from
our results as the distance at which cells transition from below
$50\%$ to above $50\%$ HII fractional density. Assuming a spherical
HII region, we compute this radius as 
$r_S = \left(8 \frac{3 V}{4\pi}\right)^{1/3}$, where $V$ is the volume
comprised of all ionized cells (i.e. where $\mn_{HII}/\mn_H \ge 0.5$), and
the additional factor of 8 arises due to the fact that our source is
in the corner, so we must mirror $V$ into the other 7 octants.  
We also plot the error in the computed I-front
radius and velocity for varying mesh sizes.  As can be clearly seen,
the computed I-front position is highly accurate, even for coarse
spatial grids, with the corresponding accuracy increasing as the mesh
is resolved.  

\begin{figure}
\centering
\parbox{\columnwidth}{
\hspace{-0.7cm}
\subfigure{\includegraphics[width=0.6\columnwidth]{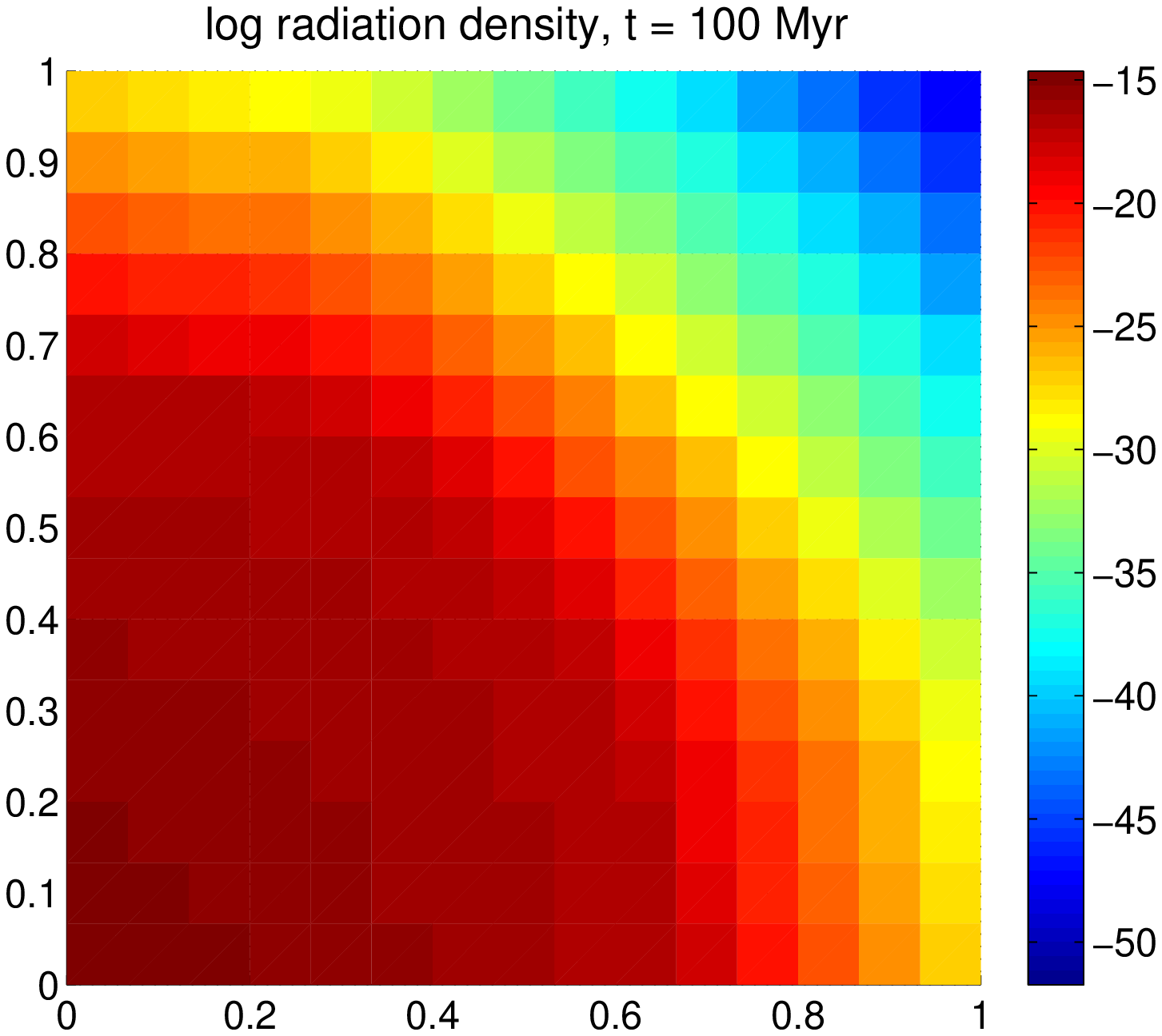}}
\hspace{-0.9cm}
\subfigure{\includegraphics[width=0.6\columnwidth]{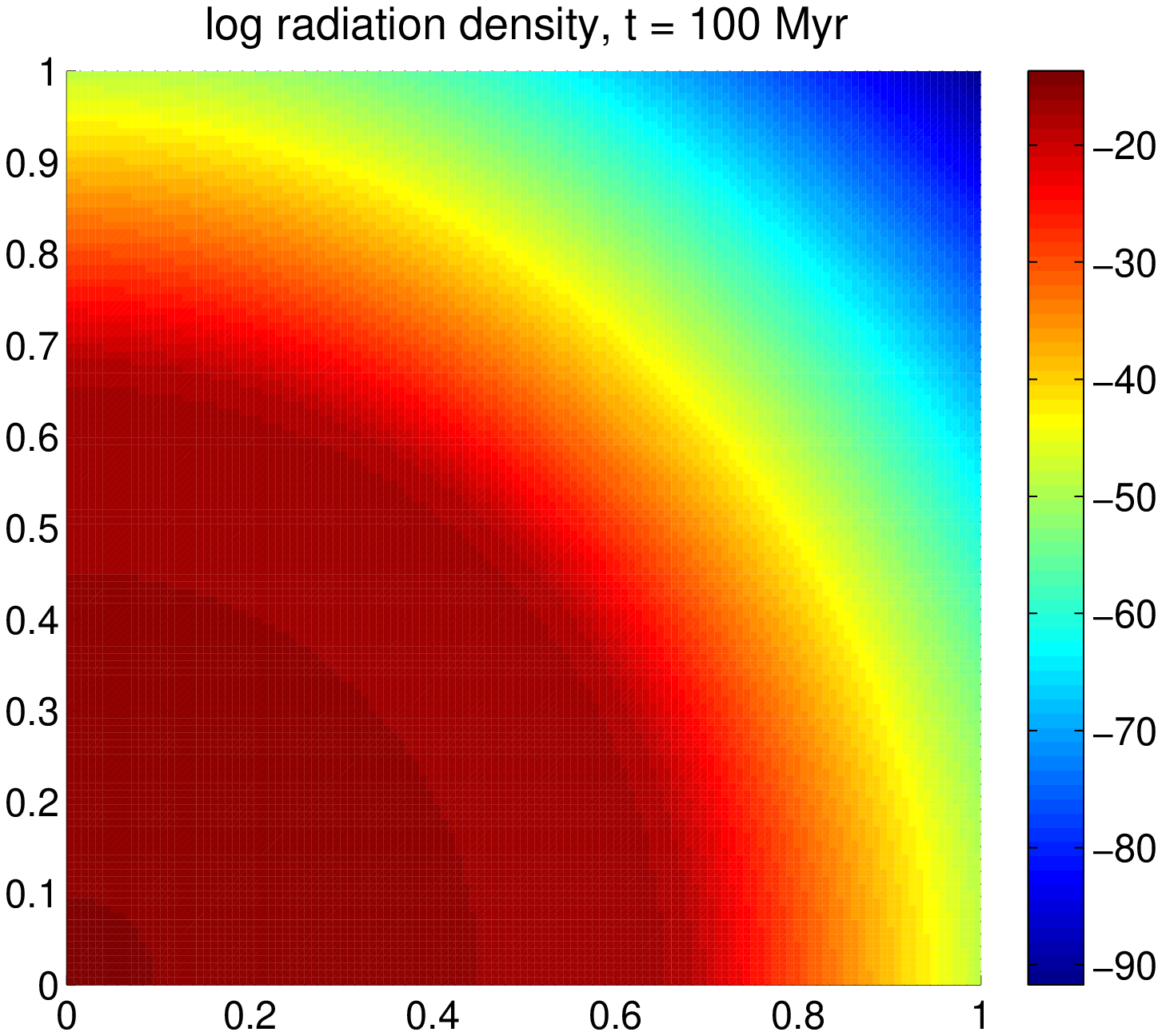}}
}
  \caption{Cross-sections of the radiation density through the
    ionization source at $t$=100 Myr: $16^3$ grid (left) and $128^3$ grid (right).
    Note the convergence to the spherical analytical shape.}  
  \label{fig:IlievEtAl1_crosssections}
\end{figure}

In Figure \ref{fig:IlievEtAl1_crosssections} we show cross-sections of
the radiation energy density through the ionization source for $16^3$
and $128^3$ grids.  We note that although the spherical front is
jagged for coarse grids, as the mesh is refined we approach the
physically-accurate spherical profile.  Moreover, this demonstrates that
although the flux limiter (\ref{eq:limiter}) is based on
one-dimensional derivatives, it does not result in anisotropic
propagation biased along axial directions.

\subsection{Cosmological Radiative Ionization}
\label{sec:sg_test}

We now perform the same test as above, but within a
cosmologically-expanding universe.  The problem is originally due to
Shapiro \& Giroux \cite{ShapiroGiroux1987}, and combines cosmology,
radiative transfer and chemical ionization.  Here, the physics of
interest is again the expansion of a HII region in uniform gas around
a single monochromatic ($h\nu = 13.6$ eV) ionizing source. Again, the
ionization front should propagate quickly at first, approaching the
Str{\"o}mgren radius, but then should begin to lag behind as
cosmological expansion drives the Str{\"o}mgren radius outward faster
than the I-front can propagate.  Due to the cosmological expansion,
the Str{\"o}mgren radius changes in time, and is given by
\begin{equation}
\label{eq:SG-stromgren_radius}
   r_S(t) = \left[\frac{3\dot{N}_{\gamma}}{4\pi \alpha_B
      \mn_H(t)^2}\right]^{1/3},
\end{equation}
where although the Hydrogen number density $\mn_H$ is spatially static,
it diminishes due to cosmological expansion by a factor of
$a^{-3}(t)$.  Defining the parameter 
$\lambda = \alpha_B\mn_{H,i} / H_0 / (1+z_i)$, where the subscript $i$
refers to the quantity at the initial redshift $z_0$, the analytical
solution is given by 
\begin{align}
  \label{eq:SG-solution}
  r_I(t) &= r_{S,i} \left(\lambda\,e^{-\tau(t)}\int_{1}^{a(t)}
    e^{\tau(\tilde{a})}\left[1-2q_0+2q_0(1+z_i)/\tilde{a}\right]^{-1/2}
    \mathrm{d}\tilde{a}\right)^{1/3}, \\
  \notag \text{where}\qquad&\\
  \label{eq:SG-tau}
  \tau(a) &= \lambda\,\left[6\,q_0^2\,(1+z_i)^2\right]^{-1}\left[F(a)-F(1)\right], \\
  \label{eq:SG-F}
  F(a) &= \left[2 - 4q_0 - 2q_0(1+z_i)/a\right]
  \left[1-2q_0 + 2q_0(1+z_i)/a\right]^{1/2}. 
\end{align}
Here, $q_0$ is the cosmological deceleration parameter and $z_i$ is
the initial redshift. We perform four of the tests provided in the
original paper \cite{ShapiroGiroux1987}:  $q_0$ of 0.5 and 0.05, and
$z_i$ of 4 and 10.  These correspond to the parameters:
\begin{itemize}
\item $(q_0,z_i)=(0.5,4)$: $L_i\approx80$ kpc, $H_0=0.5$,
  $\Omega_m=1.0$, $\Omega_{\Lambda}=0$, $\Omega_b=0.2$
\item $(q_0,z_i)=(0.05,4)$: $L_i\approx60$ kpc, $H_0=1$,
  $\Omega_m=0.1$, $\Omega_{\Lambda}=0$, $\Omega_b=0.1$
\item $(q_0,z_i)=(0.5,10)$: $L_i\approx36$ kpc, $H_0=0.5$,
  $\Omega_m=1.0$, $\Omega_{\Lambda}=0$, $\Omega_b=0.2$
\item $(q_0,z_i)=(0.05,10)$: $L_i\approx27$ kpc, $H_0=1$,
  $\Omega_m=0.1$, $\Omega_{\Lambda}=0$, $\Omega_b=0.1$
\end{itemize}
where $L_i$ is the initial box size, $H_0$ is the Hubble constant,
$\Omega_m$ is the contribution of all non-relativistic matter to the
gas energy density at $z=0$, in units of the value required to close
the universe, similarly $\Omega_{\Lambda}$ and $\Omega_b$ are the
contributions of the cosmological constant and the baryonic matter to
the energy density, respectively.  These two types of cosmology result
in slightly different functions for the expansion coefficient $a$.
For the case of $q_0=0.05$, this value comes from equqations (13-3)
and (13-10) in \cite{Peebles1993}.  For the case $q_0=0.5$, we use the
standard formula $a = (1+z)^{-1}$.  We begin all problems with an
initial radiation energy density of $E = 10^{-35}$ erg cm$^{-3}$ and an
initial ionization fraction (HII/H) of 0. 
The initial density is dependent on $q_0$, with
$\rho_{b,i}=1.175\times10^{-28}$ g cm$^{-3}$ for $q_0=0.5$, and
$\rho_{b,i}=2.35\times10^{-28}$ g cm$^{-3}$ for $q_0=0.05$.
All simulations are run from the initial redshift $z_i$ to $z=0$.
All other problem parameters are identical to those in
\S\ref{sec:Iliev1_test}.  All implicit solver parameters are also
identical to those in \S\ref{sec:Iliev1_test}, but with desired
solution accuracy $\tau_{\text{tol}}=0.001$ and inexactness parameter
$\delta_k=10^{-13}\|f(U_k)\|$. 

\begin{figure}
 \centerline{
   \includegraphics[width=0.5\linewidth]{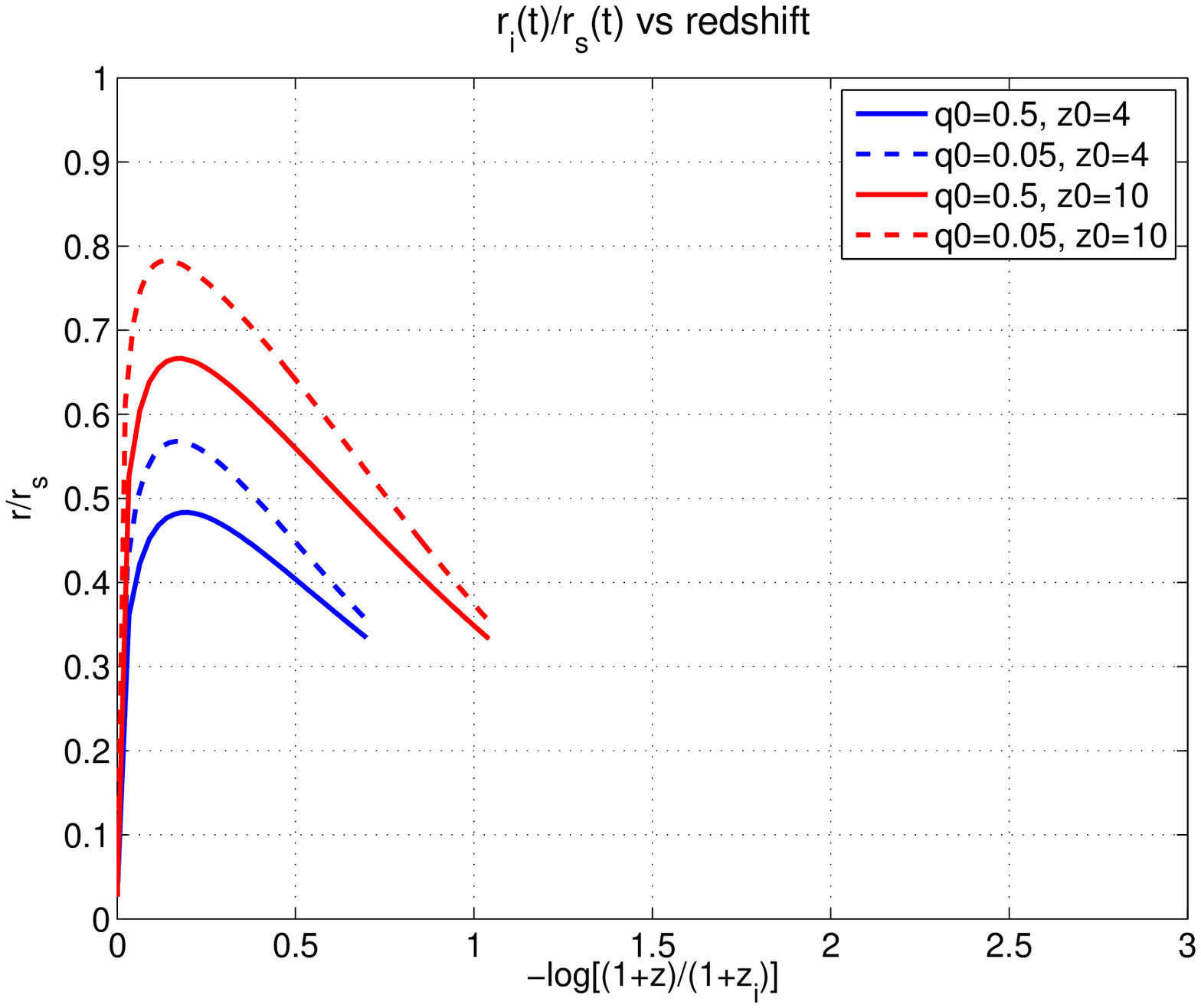}   
   \hfill
   \includegraphics[width=0.5\linewidth]{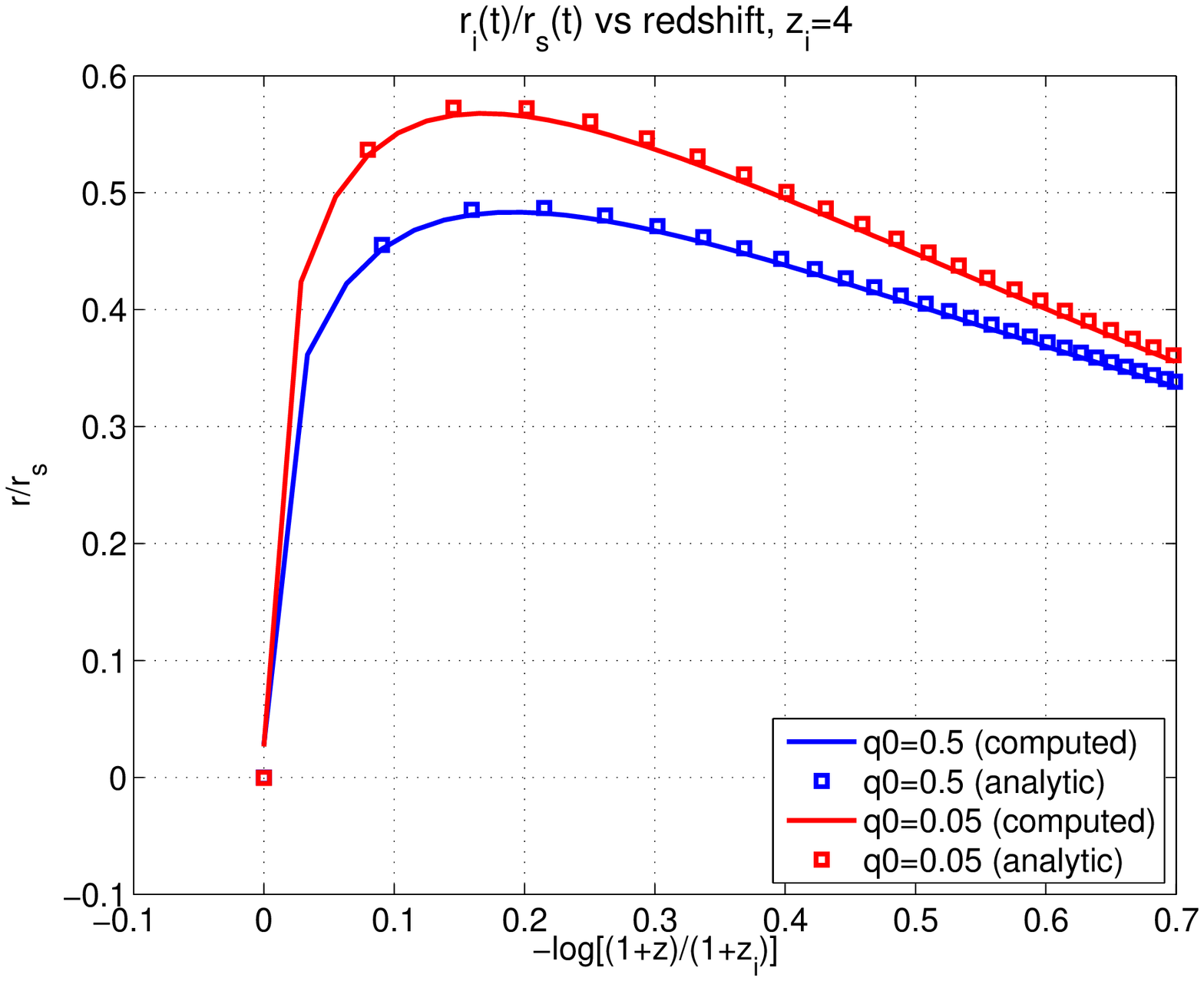}}
 \caption{Left: I-front radii vs. redshift for the four tests,
   $q0=0.5$ and 0.05, and $z_i=4$ and 10.  Right: I-front radii vs
   redshift for the $z_i=4$ tests; analytical solution values
   are given by the open squares.}
 \label{fig:ShapiroGiroux_evolution}
\end{figure}

\begin{figure}
 \centerline{
   \includegraphics[width=0.5\linewidth]{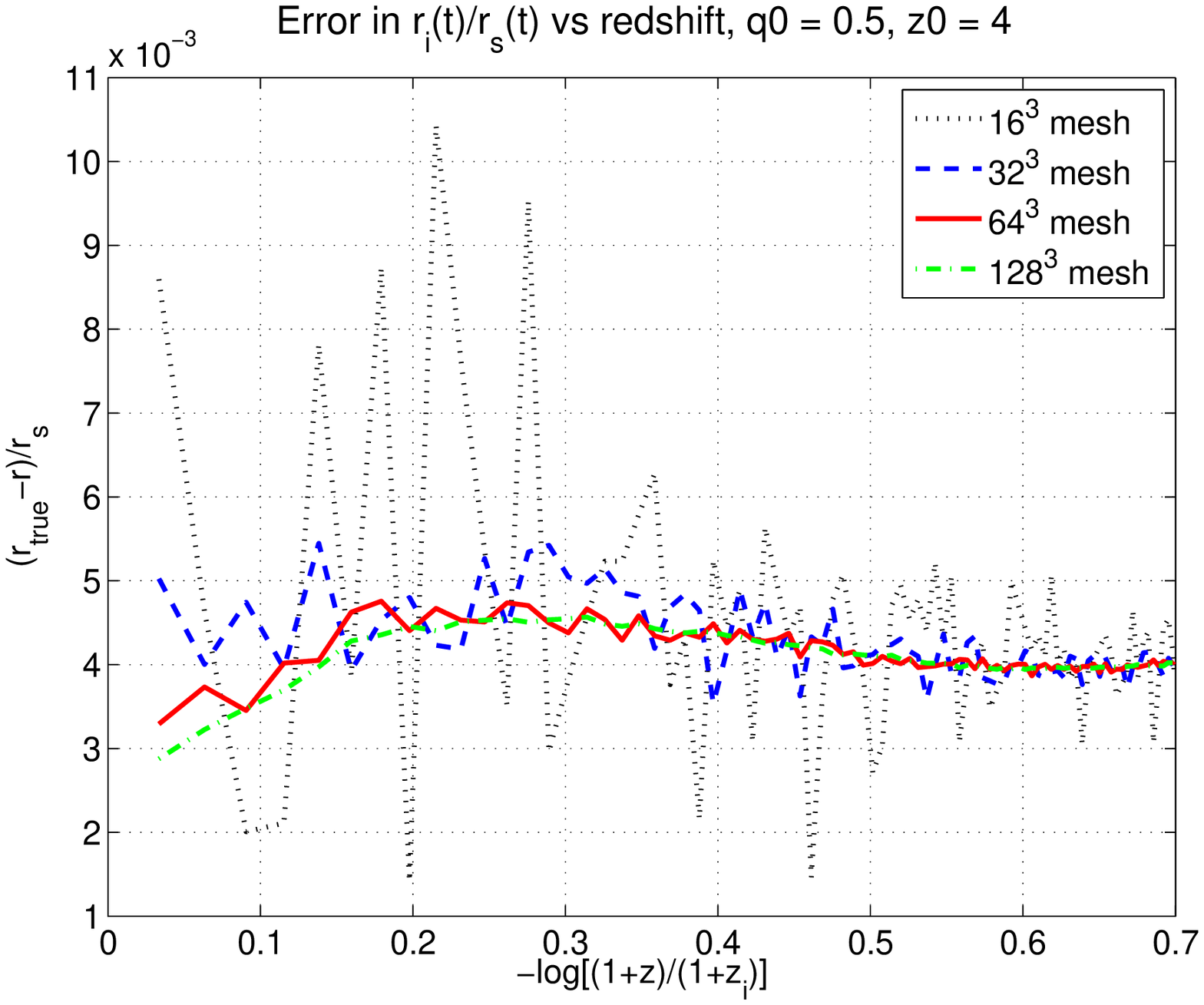}
   \hfill
   \includegraphics[width=0.5\linewidth]{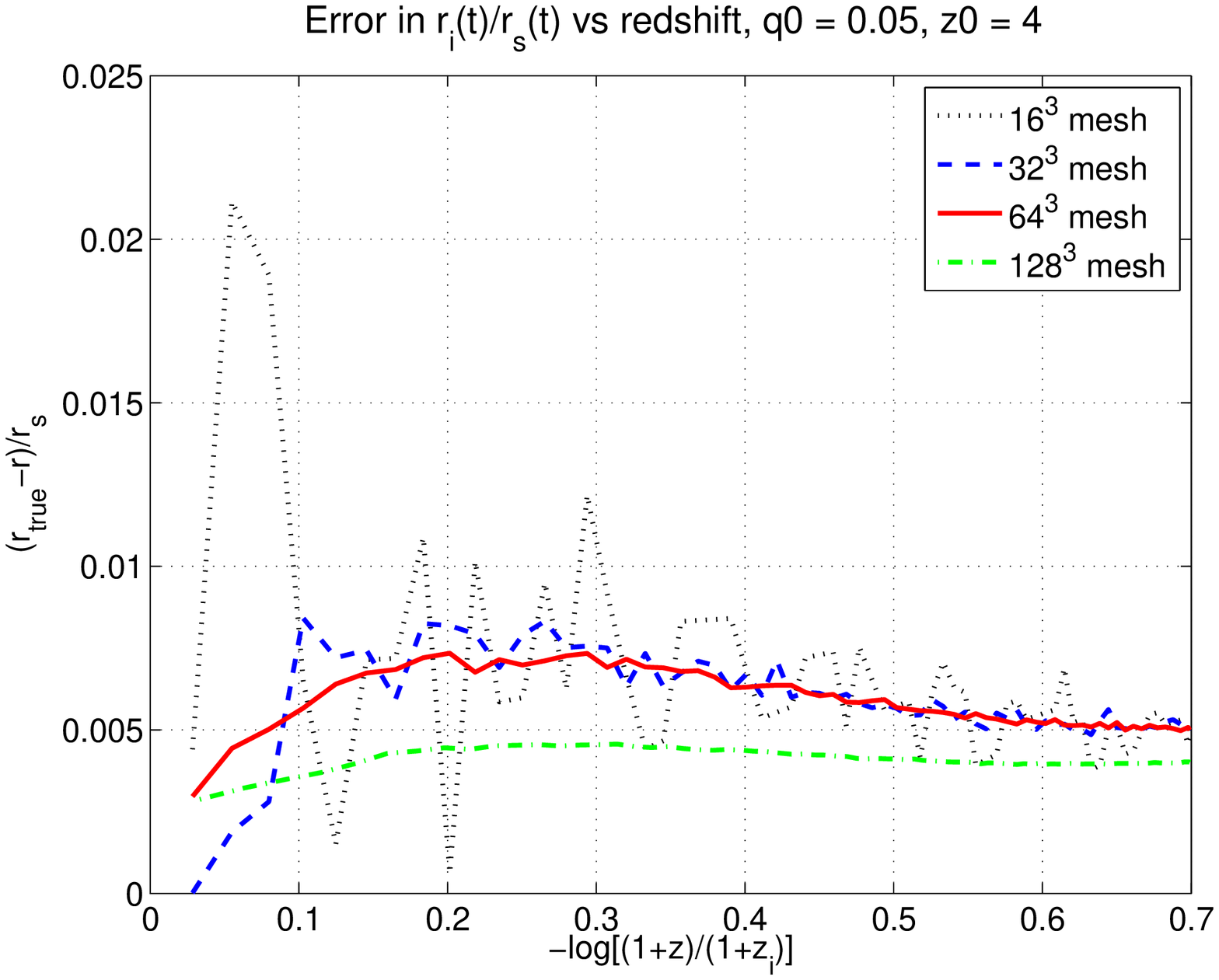}}
 \caption{Convergence of I-front radius vs. redshift for the two cases
   $q_0=\{0.5,0.05\}$ and $z_i=4$ as the mesh is refined: spatial
   meshes shown are $16^3$ (black dotted), $32^3$ (blue dashed),
   $64^3$ (red solid), and $128^3$ (green dot-dashed).} 
 \label{fig:ShapiroGiroux_convergence}
\end{figure}

In Figure \ref{fig:ShapiroGiroux_evolution} we plot the
scaled, spherically-averaged I-front position with respect to scaled
redshift for each of the four tests (with axes identical to
\cite{ShapiroGiroux1987}, Figure 1a), as well as the corresponding
plots for just the $z_i=4$ tests along with their analytical
solutions.  These solutions all used a uniform $128^3$ spatial mesh.
In Figure \ref{fig:ShapiroGiroux_convergence} we plot the error in the
computed I-front radius for varying mesh sizes for the two cases of
$q_0=0.5$ and $q_0=0.05$ with $z_i=4$.  Again, the accuracy in the
computed I-front position improves as the mesh is resolved.

\subsection{Hydrodynamic Radiative Ionization}
\label{sec:wn_test}

We now incorporate hydrodynamic motion into the mixture of physical
processes, and examine a problem due to Whalen \& Norman
\cite{WhalenNorman2006} that combines radiation, hydrodynamics and
chemical ionization (but not cosmology).   The problem is nearly
identical to that from \S\ref{sec:Iliev1_test}, but now in a dynamic
medium (varying temperature, density and velocity).  Again, the
physics of interest is the expansion of a HII region in an
initially uniform gas around a single ionizing source, though now the
source emits $\dot{N}_{\gamma} = 5\times10^{48}$ photons per second
with a frequency profile given by a $T_B=10^5$  blackbody spectrum.
Here, the ionization front should propagate quickly at first, slowing
until it reaches the Str{\"o}mgren radius (\ref{eq:Stromgren_radius}), 
at which point the I-front transitions from radiation-driven (R-type)
to dynamically-driven (D-type), and the high pressure of the ionized
and heated gas inside the HII region continues to push the I-front out
past the Str{\"o}mgren sphere.  The expansion will finally stall when
a pressure equilibrium has been reached, at a radius $r_f =
\left(2T_i/T_e\right)^{2/3}r_S$, where $T_i$ is the temperature of the
ionized gas behind the front, and $T_e$ is the temperature of the
ionized gas ahead of the front.  Analytical models for the initial
radiation-only phase predict 
\begin{equation}
\label{eq:Ifront_radiation_phase}
  r^R_I = r_S \left[1-e^{-t/t_{rec}}\right]^{1/3},
\end{equation}
where $t_{rec} = \left[\alpha_B(T_i) \mn_H\right]^{-1}$ is the
recombination time (assumed constant in this phase).  Analytical
models for the subsequent pressure-only phase predict 
\begin{equation}
\label{eq:Ifront_pressure_phase}
  r^p_I = r_S\left(1 + \frac{7\,c_s\,t}{4\,r_S}\right)^{4/7},
\end{equation}
where $c_s= \sqrt{p_I/\rho_I}$ is the sound speed in the ionized gas.
We note that due to the dynamic nature of this problem, the
true solution should lie between the two curves
(\ref{eq:Ifront_radiation_phase}) and
(\ref{eq:Ifront_pressure_phase}), since both radiation and gas
pressure play a role throughout the dynamics, and neither $T_i$ or
$T_e$ are in fact constant behind or beyond the I-front.  

We use the following problem parameters: the initial gas temperature
is set to $T=10^2$ K; the initial radiation energy density is $E^0 =
10^{-20}$ erg cm$^{-3}$; the hydrogen is initially fully neutral
(i.e.~HII/H=0); the spatial domain is a 15 kpc box.  We run for a
simulation time of 1 Gyr, which is not long enough to
reach final the final equilibrium $r_f$, but well past the transition
from R-type to D-type.  The implicit solver parameters are identical
to those in \S\ref{sec:Iliev1_test}, but with a linear solver
parameter $\delta_k=10^{-9}\|f(U_k)\|$.

\begin{figure}
\centering
\parbox{\columnwidth}{
\subfigure{\includegraphics[width=0.5\columnwidth]{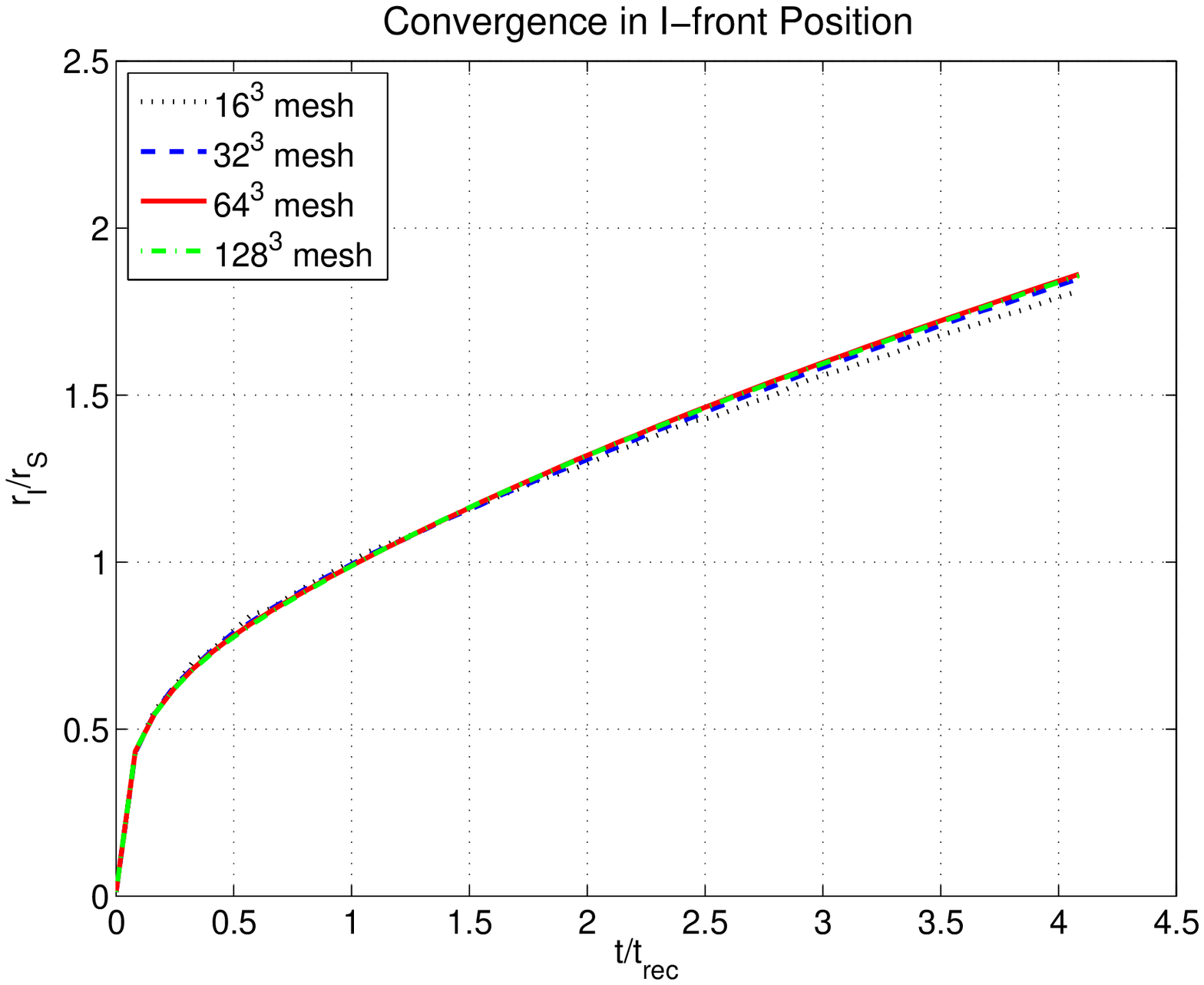}}
\hfill
\subfigure{\includegraphics[width=0.5\columnwidth]{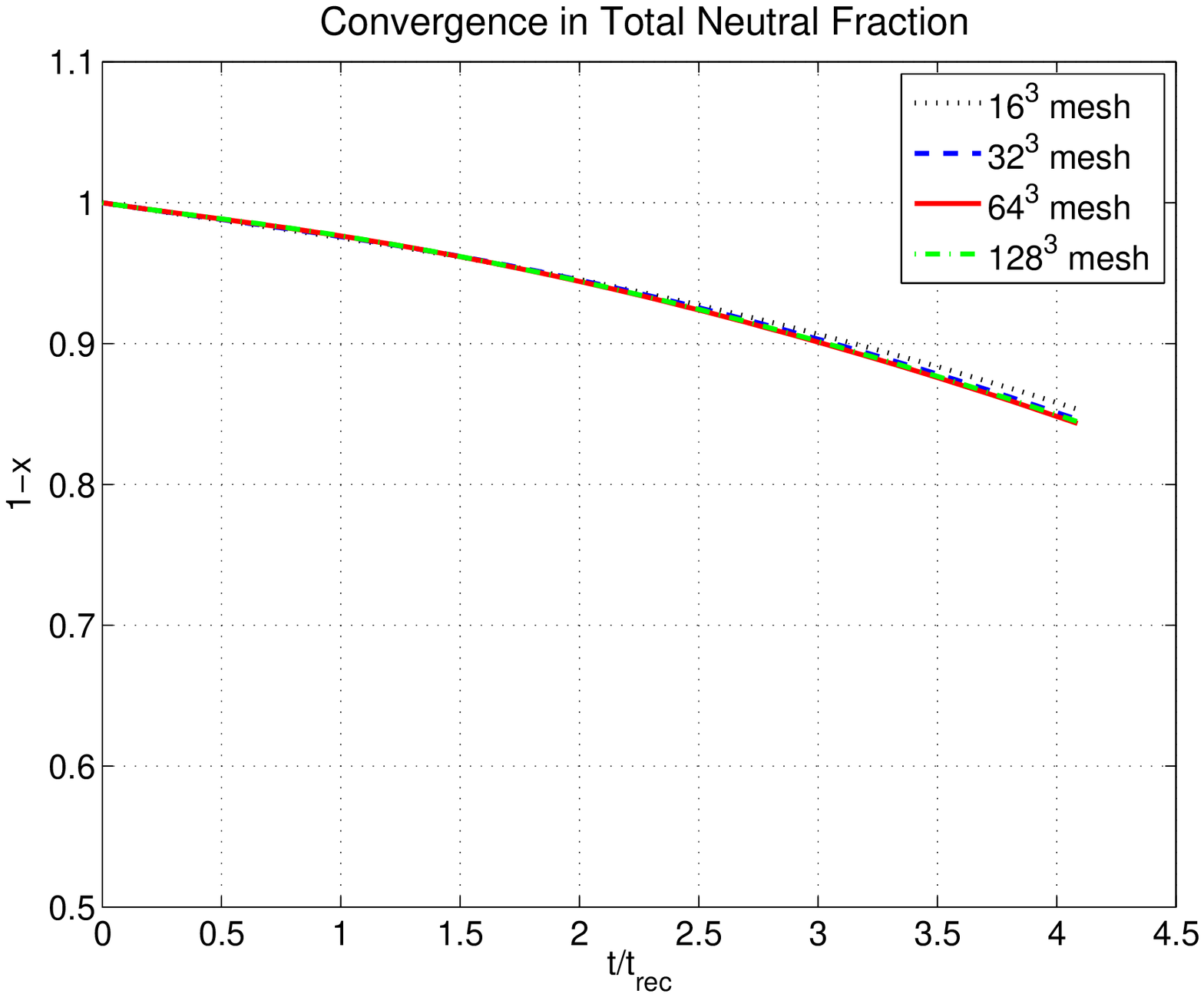}}
}
\parbox{\columnwidth}{
\hspace{-0.25cm}
\subfigure{\includegraphics[width=0.525\columnwidth]{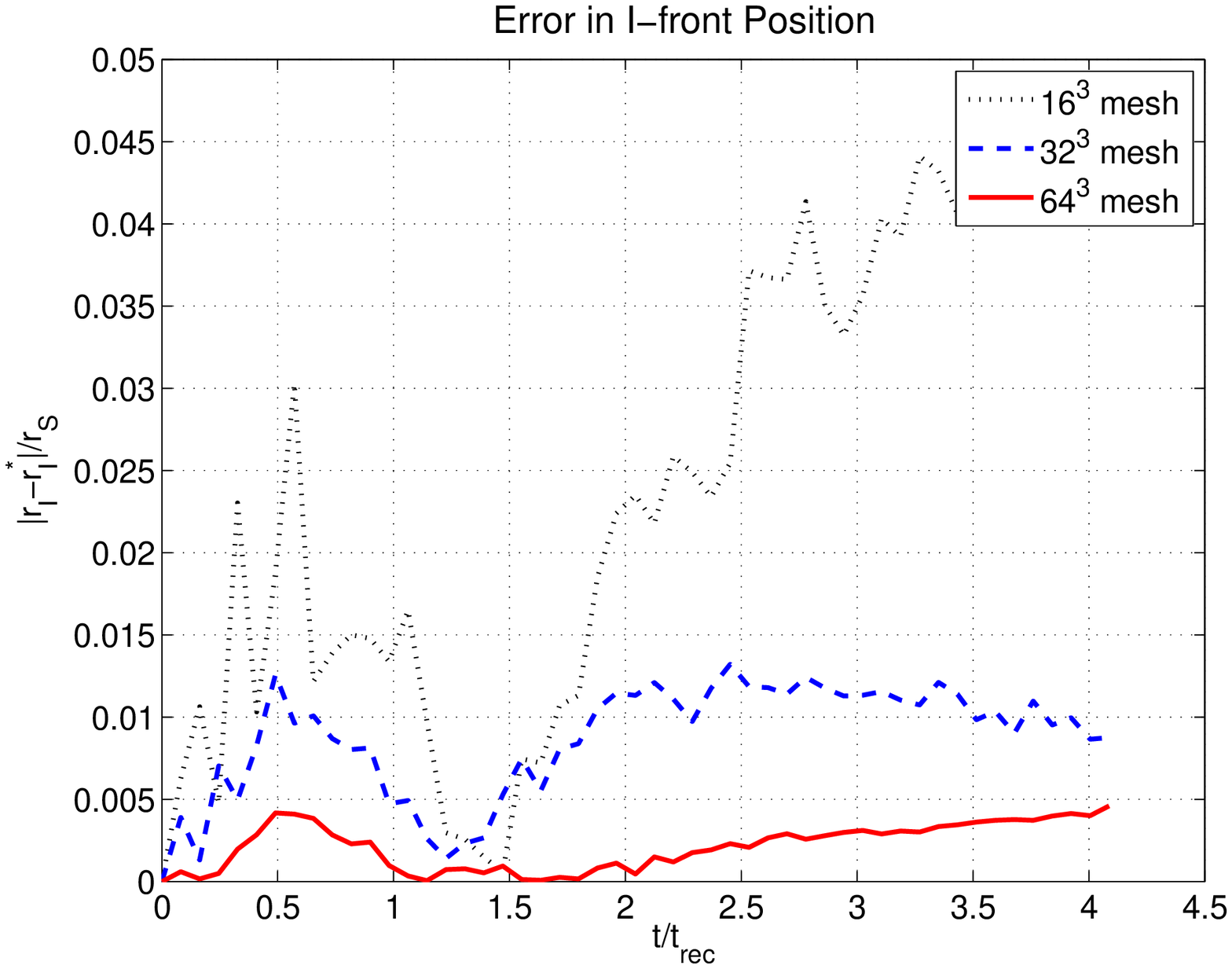}}
\hfill
\subfigure{\includegraphics[width=0.5\columnwidth]{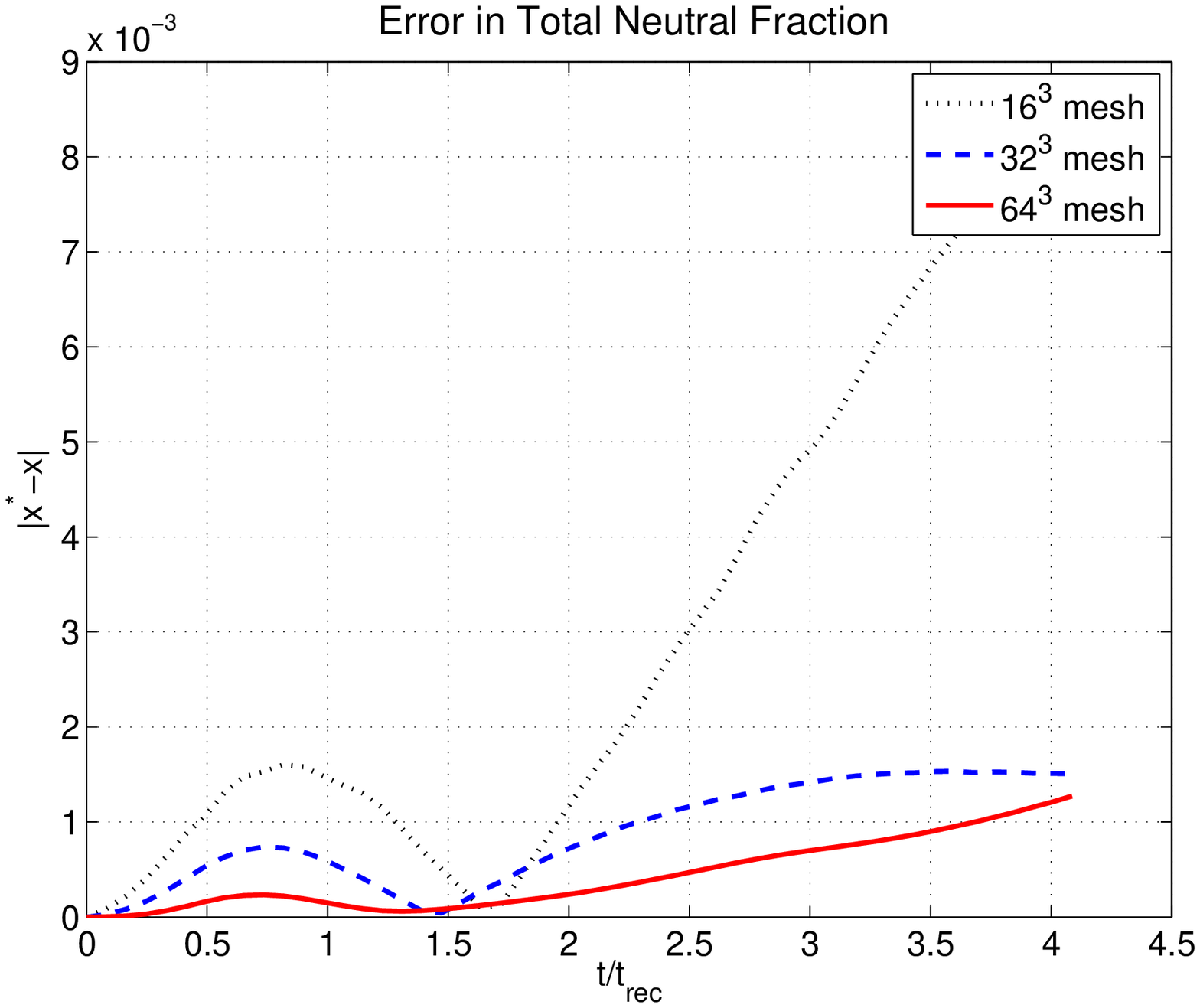}}
}
 \caption{Convergence of the hydrodynamic ionization test with mesh
   refinement.  Left: overlay of computed I-front position (top) and
   error (bottom) for varying mesh sizes.  Right: overlay of
   computed neutral fraction (top) and error for varying grids.}
 \label{fig:WhalenNorman-convergence}
\end{figure}

Results from these tests are shown in Figure
\ref{fig:WhalenNorman-convergence}, which plot the computed I-front
position and neutral fractions for various spatial meshes, along with
the ``error'' in these quantities.  Since this test problem does not
have true analytical solutions, we compute the ``error'' as the
deviation in each solution from the most-refined $128^3$ mesh
solution. 

\begin{figure}
 \centerline{
   \hspace{-0.4cm}
   \includegraphics[width=0.55\linewidth]{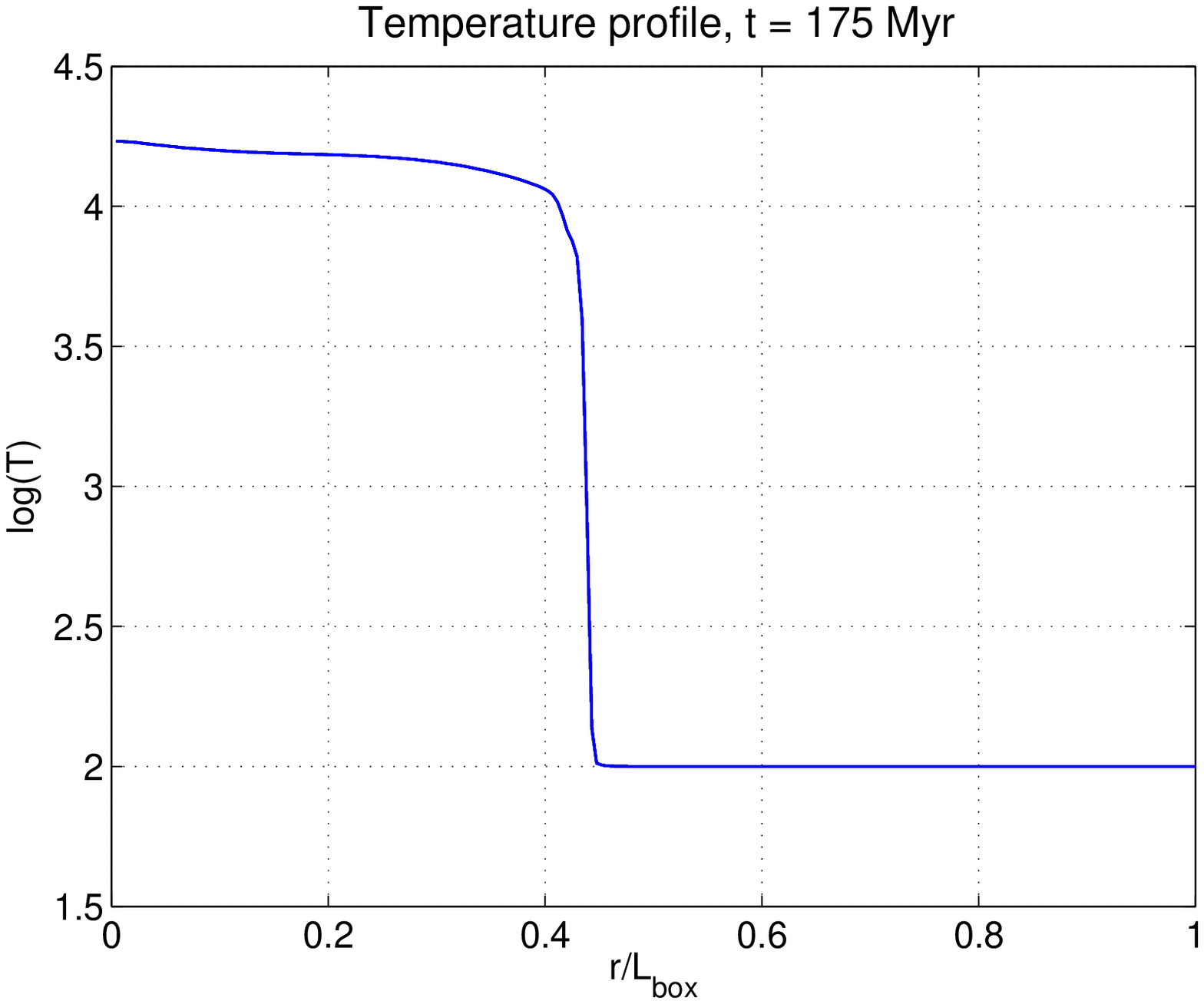}
   \hspace{-0.4cm}
   \includegraphics[width=0.55\linewidth]{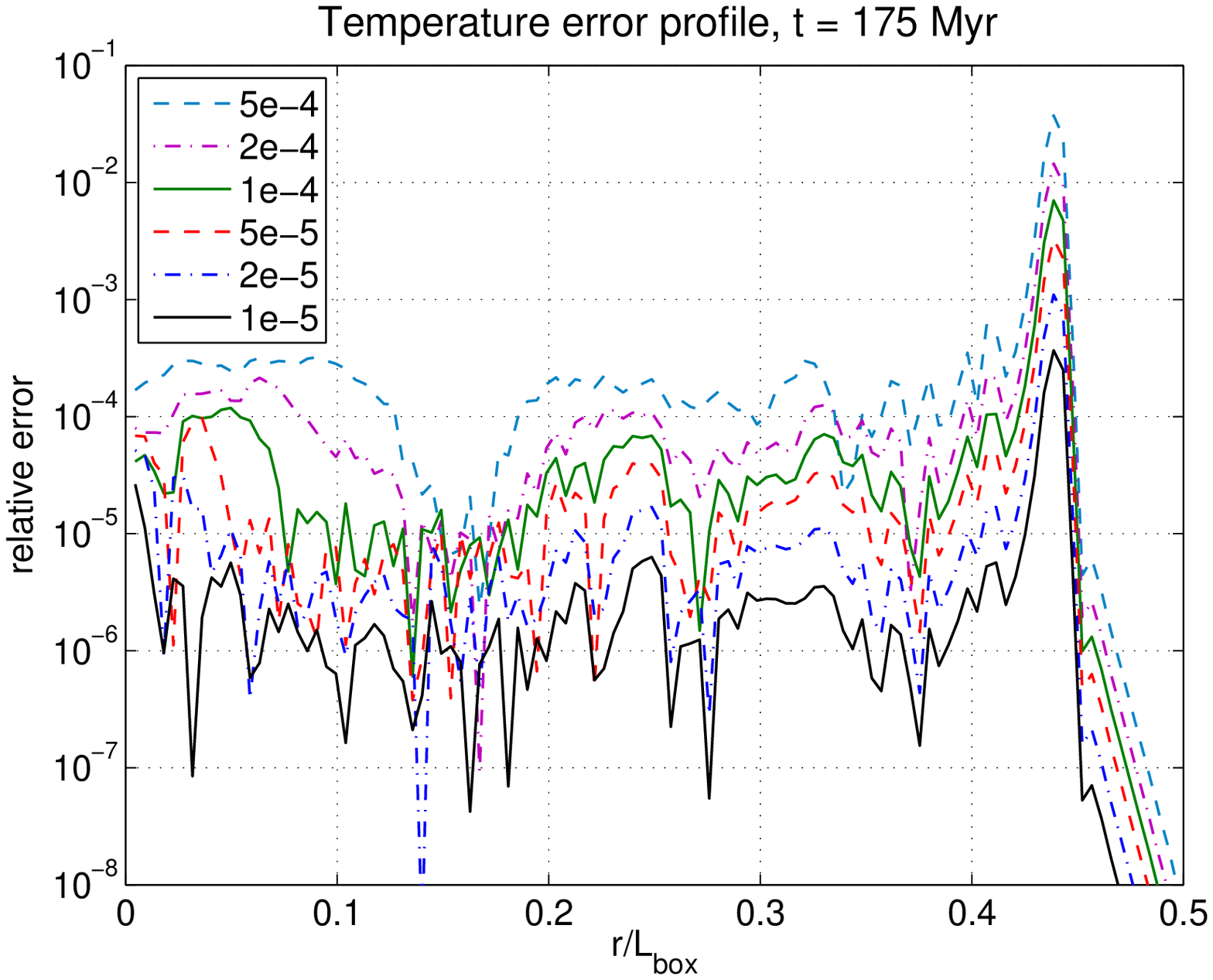}}
 \caption{Convergence of temperature profile as the time step size is 
   refined.  Left: the spherically-averaged temperature profile at 175
   Myr.  Right: relative errors in temperature for varying time step
   sizes.  Average relative errors were 2.6e-4, 1.0e-4,
   4.9e-5, 2.4e-5, 8.2e-6, and 2.7e-6, for the coarsest to finest time
   steps, respectively.}
 \label{fig:WhalenNorman-dt_convergence}
\end{figure}

We also use this problem to examine the effect of our operator split
solution strategy on the temporal accuracy of the solver.  In Figure 
\ref{fig:WhalenNorman-dt_convergence} we plot the spherically-averaged
temperature profile for a $128^3$ spatial grid at 175 Myr, and the
associated relative errors found through varying the time step
size. The error plot has been zoomed in around the heated region and
front.  We note that although both PPM and the described implicit
sub-solver are both up to second-order accurate, the splitting reduces
the resulting accuracy to slightly better than first-order in time.
We also note that the adaptive time-stepping strategy from Section
\ref{subsec:adaptive_time_stepping} results in average time steps of
6.0e-4, 5.7e-4, 4.0e-4 and 2.7e-4 for the $16^3$ through $128^3$
grids, respectively; whereas the hydrodynamic CFL-limited time steps
for these same grids are 3.1e-3, 1.4e-3, 6.0e-4 and 2.8e-4.  Hence the
radiation and ionization dynamics drive the system for coarser meshes,
while at finer mesh sizes the hydrodynamic CFL condition begins to
dominate.

\subsection{Weak Scaling}
\label{sec:scaling_tests}

As described in our introduction and throughout the description of our
numerical methods, a key goal in introducting a fully implicit
solution mechanism for the stiff components in radiation,
hydrodynamics and chemical ionization simulations is the eventual
scalability of such a solver to very large problem sizes.  We
therefore investigate the weak scaling of the implicit solver on the
cosmological radiative ionization problem from section
\ref{sec:sg_test}.  For these tests, we emulate the setup from the
$q_0=0.5$ and $z_i=4$ test, but here we place an ionizing source in
the center of {\em each processor's} subgrid.  Moreover, for these
weak scaling tests we increase the domain size and mesh size
proportionately to the number of processors, where each processor owns
a $64^3$ grid and an initial subgrid box size of 80 kpc.  We then
run problems that scale up from 1 to 4096 processors, resulting in
spatial grids ranging from $64^3$ to $1024^3$.  Moreover, since we are
investigating the scaling properties of the numerical methods, we
shorten the simulation time to evolve from $z_i=4$ to $z=3$ in order
to conserve on supercomputer resources, while retaining the portion of
the simulation with the most rapidly-evolving dynamics.  All runs were
performed on 
the NSF Kraken machine (using 2 cores/node). 

\begin{figure}
 \centerline{
   \includegraphics[width=0.7\linewidth]{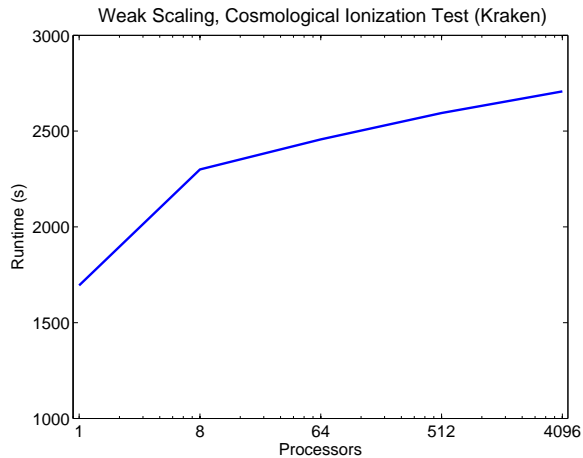}}
 \caption{Weak scaling results for the cosmological HII-region
   expansion test.}
 \label{fig:weak_scaling}
\end{figure}

We show the runtimes associated with these tests in Figure
\ref{fig:weak_scaling}, and provide detailed statistics from each 
run
in Table \ref{table:scaling_stats}.  We note that on this
architecture, the solver demonstrates near-perfect scalability,
with modest increases in runtime for parallel versus serial runs, and
only marginal increases in solution time as the parallelism is
increased.  The reason for this is the near-constant number of
iterations required by the nonlinear Newton solver, the outer CG
linear solver, and the inner multigrid preconditioner.  Therefore the
increase in run time may be directly attributed to the ideal $\mathcal
O(\log p)$ increase in runtimes typical of multigrid methods, allowing
near-optimal scalability to the limits of modern supercomputer
resources. 

\begin{table}
{\scriptsize
  \centering
  \begin{tabular}{|c|c|c|c|c|c|c|} \hline
     Mesh & Processors & Time Steps & Run Time & Newton Its & CG
     Its & MG V-cycles \\
     \hline \hline
     $64^3$   &    1  & 266 & 1694.38 & 322 & 914 & 2991 \\
     $128^3$  &    8  & 265 & 2299.60 & 274 & 799 & 2575 \\
     $256^3$  &   64  & 265 & 2456.58 & 268 & 787 & 2524 \\
     $512^3$  &  512  & 264 & 2594.50 & 265 & 780 & 2510 \\
     $1024^3$ & 4096  & 264 & 2707.30 & 265 & 780 & 2510 \\
     \hline
  \end{tabular}
  \caption{Cosmological Weak Scaling Statistics.}
  \label{table:scaling_stats}
}
\end{table}

%% file: conclusions.tex
\section{Conclusions}
\label{sec:conclusions}

We have described an implicit formulation for coupling cosmological
radiation transport, chemical ionization and gas energy feedback
within Enzo hydrodynamics simulations.  The formulation is based on an
operator-splitting between the non-stiff hydrodynamics and stiff 
radiation-ionization-energy feedback physical processes, in which the
stiff processes are solved within a fully-implicit
Newton-Schur-Krylov-Multigrid framework.  

Through numerous tests, we have demonstrated that this solver is
accurate and stable, allowing simulations of a wide variety of
physical environments from the laboratory scale to the astronomical and
even cosmological scales.  Moreover, through the choice of numerical
methods that form the implicit solver, it demonstrates ideal
scalability for such coupled physics simulations.  In addition, this
implicit formulation is highly extensible, and may easily be adjusted
to allow new physical processes such as magnetic fields,
multi-frequency radiation transfer, and additional chemical species.
Finally, we are in the process of extending this approach to allow for
adaptive spatial discretizations (AMR), which should require
adjustments to only the inner multigrid linear solver.

%% file: acknowledgements.tex
\section*{Acknowledgements}
\label{sec:acknowledgements}

The authors wish to acknowledge the many insightful discussions with
Frank Graziani, Louis Howell, Carol Woodward, and Doug Swesty during 
the course of this work; as well as Rob Lowrie for providing
semi-analytical solutions to (\S\ref{sec:le_test}).  The verification
tests were performed on the Itanium 2 linux cluster {\em Thunder} at
the Lawrence Livermore National Laboratory, and on the IBM Power 4
supercomputer {\em DataStar} at the San Diego Supercomputing Center.
The weak scaling tests were performed on the Cray XT4 system 
{\em Kraken} at the National Institute for Computational Sciences.
This work was supported in part by NSF grants AST-0708960 and
AST-0808184, NASA ATFP grant NNX08AH26G, and by UCSD-LLNL
collaborative grant ``LUSciD: Scientific Data Management'' to MLN, 
and performed in part under the auspices of the U.S. Department of Energy
by the University of California, Lawrence Livermore National
Laboratory under Contract W-7405-ENG-48.